\documentclass{aa}  

\usepackage{graphicx}
\usepackage{txfonts}
\usepackage{amsmath}
\usepackage{tabularray}
\usepackage{ulem}
\usepackage{float}
\usepackage{stfloats}
\usepackage{multicol}
\usepackage{color,soul}
\definecolor{lightblue}{rgb}{.70,.95,1}
\sethlcolor{lightblue}

\newcommand{\kpc}{~\mathrm{kpc}}
\newcommand{\pc}{~\mathrm{pc}}
\newcommand{\kms}{~\mathrm{kms}^{-1}}
\newcommand{\kmskpc}
{~\mathrm{kms}^{-1}\mathrm{kpc}^{-1}}
\newcommand{\bsym}[1]{\boldsymbol{#1}}
\newcommand{\bx}{\bsym{x}}
\newcommand{\bv}{\bsym{V}}
\newcommand{\bJ}{\bsym{J}}
\newcommand{\btheta}{\bsym{\theta}}

\begin{document}

   \title{A non-axisymmetric potential for the Milky Way disk}
   
   \author{Y. R. Khalil
          \inst{1}
          {\thanks{\email{yassin.khalil@unistra.fr}}}
          \and
          B.~Famaey\inst{1}
          \and
          G.~Monari\inst{1}
          \and
          M.~Bernet\inst{2,3,4}
          \and
          A.~Siebert\inst{1}
          \and
          R.~Ibata\inst{1}
          \and
          G. F.~Thomas\inst{5,6}
          \and
          \\ P. Ramos\inst{7}
          \and
          T.~Antoja\inst{2,3,4}
          \and
          C.~Li\inst{8,9}
          \and
          S.~Rozier\inst{10}
          \and
          M.~Romero-G\'omez\inst{2,3,4}
          }

   \institute{Universit\'e de Strasbourg, CNRS  UMR  7550,  Observatoire  astronomique  de  Strasbourg, 11 rue de l'Université, 67000  Strasbourg,  France
              \and
              Departament de Física Quàntica i Astrofísica (FQA), Universitat de Barcelona (UB), C Martí i Franquès, 1, 08028 Barcelona, Spain 
              \and
              Institut de Ciències del Cosmos (ICCUB), Universitat de Barcelona (UB), C Martí i Franquès, 1, 08028 Barcelona, Spain
         \and
         Institut d’Estudis Espacials de Catalunya (IEEC), Edifici RDIT, Campus UPC, 08860 Castelldefels, Barcelona, Spain
         \and
             Instituto de Astrof\'isica de Canarias, E-38205 La Laguna, Tenerife, Spain 
        \and 
             Universidad de La Laguna, Dpto. Astrof\'isica, E-38206 La Laguna, Tenerife, Spain
             \and
             National Astronomical Observatory of Japan, Mitaka-shi, Tokyo 181-8588, Japan
             \and
             School of Astronomy and Space Science, Nanjing University, Nanjing 210093, People's Republic of China
             \and
             Key Laboratory of Modern Astronomy and Astrophysics (Nanjing University), Ministry of Education, Nanjing 210093, People's Republic of China
             \and
             School of Mathematics and Maxwell Institute for Mathematical Sciences, University of Edinburgh, Kings Buildings, Edinburgh, EH9 3FD, UK
             }

   \date{}

  \abstract
   {We provide a purely dynamical global map of the non-axisymmetric structure of the Milky Way disk. For this, we exploit the information contained within the in-plane motions of disk stars from Gaia DR3 to adjust a model of the Galactic potential, including a detailed parametric form for the bar and spiral arms. We explore the parameter space of the non-axisymmetric components with the backward integration method, first adjusting the bar model to selected peaks of the stellar velocity distribution in the Solar neighborhood, and then adjusting the amplitude, phase, pitch angle and pattern speed of spiral arms to the median radial velocity as a function of position within the disk. We check {\it a posteriori} that our solution also qualitatively reproduces various other features of the global non-axisymmetric phase-space distribution, including most of the moving groups and phase-space ridges despite those not being primarily used in the adjustment. This fiducial model has a bar with pattern speed $37$~km~s$^{-1}$~kpc$^{-1}$ and two spiral modes that are two-armed and three-armed respectively. The two-armed spiral mode has a $\sim 25~\%$ local contrast surface density, a low pattern speed of $13.1\kmskpc$, and matches the location of the Crux-Scutum, Local and Outer arm segments. The three-armed spiral mode has a $\sim 9~\%$ local contrast density, a slightly higher pattern speed of $16.4\kmskpc$, and matches the location of the Carina-Sagittarius and Perseus arm segments. The Galactic bar, with a higher pattern speed than both spiral modes, has recently disconnected from those two arms. The fiducial non-axisymmetric potential presented in this paper, reproducing most non-axisymmetric signatures detected in the stellar kinematics of the Milky Way disk, can henceforth be used to confidently integrate orbits within the Galactic plane.}

   \maketitle
%
%-------------------------------------------------------------------
\section{Introduction}

The Gaia mission now provides full six-dimensional phase-space information on the Milky Way (MW) disk for a larger number of stars, and over a larger volume, than ever before. The second \citep{Gaia2018}, early-third \citep{Gaia2021}, and third \citep[DR3,][]{Gaia2023a} data releases of Gaia have represented important milestones in this regard \citep{Hunt25}. Whilst stellar vertical motions have revealed a disequilibrium of the Galactic disk \citep{Antoja2018}, possibly related to a subtle interplay between external perturbations and internal non-axisymmetries \citep[e.g.,][]{Binney2018,Laporte2019,Chengdong2023,Tremaine2023,Frankel2024}, the rich information contained solely within the in-plane motions of stars \citep[e.g.,][]{Gaia2018b,Gaia2023b} has not been fully exploited yet. Indeed, these in-plane motions should -- in principle -- allow us to get a detailed dynamical mapping of the most important non-axisymmetric structures of the MW disk, namely the Galactic bar and the spiral arms \citep{LynKaj72}. However, such a detailed mapping is still lacking. This is the focus of the present study.

The existence of the MW bar was originally hypothesized from the observations of gas kinematics \citep[][]{Vaucouleurs1964,Peters1975,Gerhard1986,Binney1991} and confirmed from (near-)infrared observations \citep[e.g.,][]{Blitz91,Sellwood1993,Weiland1994,Binney1997} as well as bulge stellar kinematics \citep[e.g.,][]{Zhao1994}. It is nowadays clear that a large fraction of stars in the bulge region indeed follow a rotating, barred, boxy/peanut-shaped structure connected to an edge-on bar \citep{BlandHawthorn2016}, whose pattern speed, $\Omega_{\rm b}$, is nevertheless still subject to debate. Less than three decades ago, a consensus emerged for a pattern speed in the range $\Omega_{\rm b} \sim 50$-$60$~km~s$^{-1}$~kpc$^{-1}$ from various lines of evidence, such as hydrodynamical simulations comparing the modeled gas flow to observed Galactic CO and HI longitude-velocity diagrams \citep{Fux1999,Englmaier1999,Bissantz2003}, the \citet{Tremaine-Weinberg1984} method applied to stars in the inner Galaxy \citep{Debattista2002}, or local stellar kinematics \citep{Dehnen1999,Dehnen2000,Fux2001} positioning the Sun marginally beyond the 2:1 Outer Lindblad Resonance (OLR) of the bar. The latter argument has been supported by numerous subsequent analyses \citep[e.g.][]{Minchev2007,Quillen2011,Antoja2012,Antoja2014,Fragkoudi2019}, while on the contrary, parallel research on the density of red clump stars in the disk \citep{Wegg2015}, gas kinematics \citep{Sormani2015,Li2016}, dynamical modeling of stellar kinematics in the inner Galaxy \citep{Portail2017}, and proper motion data from the VVV survey \citep[e.g.,][]{Clarke2019}, including with the Tremaine-Weinberg method \citep{Sanders2019}, have collectively pointed to a revised pattern speed of $\Omega_{\rm b} \sim 35$-$40$~km~s$^{-1}$~kpc$^{-1}$. \citet{Perez2017} and \citet{Monari2019a} subsequently demonstrated that the Galactic model adjusted to bulge stellar kinematics by \citet{Portail2017} could effectively replicate several observed features in local velocity space \citep[see also][]{Monari2019b,Binney2020,Donghia2020,Lucchini}. Such a lower bar pattern speed is also consistent with observed overdensities in the stellar halo phase-space \citep{Dillamore2024} and with the chemistry of the disk \citep{Haywood,Khop24}. Other intermediate pattern speeds have also been proposed \citep{Hunt2018a}, and even much lower ones \citep{Horta24}, while several studies have concluded that stellar kinematics of the disk alone were not sufficient to break the degeneracy \citep{Trick2021,Trick2022,Bernet2024}. However, kinematics of stars in the bar region itself seem to have converged to $\Omega_{\rm b} \sim 35$-$40$~km~s$^{-1}$~kpc$^{-1}$, although possible sudden variations of the pattern speed are also possible \citep{Hilmi2020}. Finally, a steady decrease of the pattern speed of the bar with time has been tentatively detected \citep{Chiba2021a,Chiba2021b}, and could explain some aspects of the vertical disequilibrium of the Galactic disk \citep{Chengdong2023}, and can partially contribute in explaining the presence of metal-poor stars with prograde planar orbits in the Solar vicinity \citep{Li2023,Yuan2023}.

Concerning the location and dynamics of spiral arms of the MW, the present observational situation is even less clear than for the bar. The suggestion that the MW could host spiral arms is certainly as old as the realization that it belongs to the realm of disk galaxies, but due to extinction, it was not before the work of \citet{Morgan1952} that these arms were identified based on the distribution of HII regions, soon followed by the kinematic analysis of the HI 21-cm line by \citet{Oort1958}. Based on the distances to OB associations and HII regions, \citet{Georgelin1976} mapped a four-armed spiral pattern, which has been repeatedly confirmed with young or gaseous tracers \citep[e.g.][]{Urquhart2014}, but not with older/redder ones that should better trace the perturbations in the Galactic potential. Indeed, \citet{Drimmel2000}, \citet{Drimmel2001}, \citet{Benjamin2005} and \citet{Churchwell2009} found with near-infrared and mid-infrared tracers that the MW seemingly hosts two main spiral arms. Collecting data on HII regions and giant molecular clouds, \citet{Hou2009} showed that models of three-armed and four-armed logarithmic spirals could connect those different spiral tracers, as reviewed in \citet{ShenZheng2020}. In summary, it is no exaggeration to say that different tracers and observations are far from converging on parameters describing the positions of each spiral arm segment in the MW disk. The so-called Local arm, for example, has been found by \citet{Gaia2023b} and \citet{Poggio2021}, tracing young stars, to be more extended -- and to have an intermediary pitch angle -- compared to \citet{Vazquez2008} where this arm rather heads outwards to the Perseus arm, or to \citet{Xu2021} where the Local arm heads inwards to the Carina–Sagittarius arm.  Similar debates exist regarding the Perseus arm and Outer arm with respect to their position in the disk and pitch angle. Perhaps most strikingly, the pitch angle of the Perseus arm has been found to be $ \sim 24^{\circ}$ in \citet{Levine2006}, compatible with results of \citet{Poggio2021} or \citet{Drimmel24}, and $\sim 9^{\circ}$ in \citet{Reid2019}, meaning that the name does not actually always refer to the same observed overdensities in the Galactic plane. The situation regarding the pattern speed of spiral arms is even more confused, as its signature can also depend on their dynamical nature and origin \citep[see][for a review]{Sellwood2022}. Indeed, numerical simulations of galactic disks offer multiple perspectives on this topic, from transient corotating structures winding up and disappearing quickly \citep[e.g.,][]{Baba2013,Hunt2018b,Hunt2019}, to multiple modes persisting over a few (or even many) galactic rotations, falsely appearing as very short-lived due to superposition of modes \citep[e.g.,][]{Sellwood2014}. In the following, our modeling procedure will follow two main guidelines. The first guideline is the current consensus that, when spirals appear as modes in simulations, these are not strictly static as in the classical density wave picture \citep{LinShu1964}, but are rather made of a recurrent cycle of groove modes \citep[seeded by a depletion of circular orbits in a narrow range of angular momenta, see, e.g.][]{Derijcke2016,Derijcke2019} that live between their inner Lindblad resonance (ILR) and OLR, where they can create new grooves which set up the recurrent cycle \citep{Sellwood2014,Sellwood2019}. They can also be edge modes \citep{Fiteni}. The amplitudes of the individual modes grow and decay, but they are nevertheless genuine standing wave oscillations with fixed shape and pattern speed, detectable over a period of at least one rotation. The second guideline is that spectrograms of $N$-body simulations displaying joint bar and spiral perturbations tend to display spiral arms rotating more slowly than the bar; moreover spiral arms are never present within the corotation radius of the bar. These spirals live between their own ILR and OLR but are usually strongest between their ILR and corotation \citep{Quillen2011}. Our modeling approach will not take into account the possibility of winding spirals with time.

Keeping all the aforementioned caveats in mind, several tentative measurements of the amplitude and pattern speed of spiral arms have been made over time. Originally, \citet{Lin1969} proposed a 2-armed model with pitch angle of $6^\circ$ and pattern speed of $\Omega_{\rm s,2} \sim 13$~km~s$^{-1}$~kpc$^{-1}$ based on the systematic motion of gas and the displacement of moderately young stars in their classical density wave model. A more recent estimate based on the classical \citet{LinShu1964} density wave formalism has been made by \citet{Siebert2012} fitting the mean radial velocity map from the RAVE survey and finding a best-fit 2-armed spiral model with a contrast surface density of 14\%, a pitch angle of $10^\circ$ and a pattern speed of $\Omega_{\rm s,2} \sim 18.6$~km~s$^{-1}$~kpc$^{-1}$. This model however neglected the effect of the bar \citep[see, e.g.][]{Monari2014}. On the other hand, \citet{Amaral1997} argued for a superposition of a 2-armed and 4-armed spiral, both with a pattern speed of $\sim 20$~km~s$^{-1}$~kpc$^{-1}$ based on tracing back open clusters to their birth place. Such a procedure was recently carried out by \citet{CastroGinard2021}, finding a declining pattern speed with radius from $\sim 50$~km~s$^{-1}$~kpc$^{-1}$ for the Scutum arm segment to $\sim 17$~km~s$^{-1}$~kpc$^{-1}$ for the Perseus arm segment. Modeling the gas flow in the inner Galaxy, \citet{Bissantz2003} obtained a joint measurement of the bar and 4-armed spiral pattern speeds, with a very high pattern speed for the bar, $\Omega_{\rm b} \sim 60$~km~s$^{-1}$~kpc$^{-1}$, and a 4-armed spiral pattern speed of $\Omega_{\rm s,4} \sim 20$~km~s$^{-1}$~kpc$^{-1}$. More recently, again neglecting the bar, \citet{Eilers2020} applied a toy model of a logarithmic spiral to Gaia DR2 mean Galactocenric radial velocity field to suggest a contrast surface density of $\sim 10$\%, a pitch angle of $12^\circ$ and a fixed pattern speed of $\Omega_{\rm s,2} = 12$~km~s$^{-1}$~kpc$^{-1}$ for a 2-armed spiral corresponding to the Local and Outer arms. Such low pattern speeds had also previously been suggested by, e.g., \citet{Sellwood2010} based on the signature of a spiral ILR in local stellar kinematics ($\sim 8$~km~s$^{-1}$~kpc$^{-1}$ for a 2-armed spiral and $\sim 15$~km~s$^{-1}$~kpc$^{-1}$ for a 3-armed spiral). Regarding the amplitude, the most recent determination, based on the vertical Jeans equation, has found the Local arm to be the strongest local overdensity, with a contrast density of roughly 20\%  \citep{Widmark2024}.

Here, we attempt at fully exploiting the rich information encoded within the in-plane stellar motions in Gaia DR3 \citep{Gaia2023b} to constrain dynamically the non-axisymmetries of the Galaxy. Previous similar attempts include the more empirical approach of \citet{Khop20,Khop22}, as well as the recent work of \citet{Vislosky2024} comparing three hydrodynamical simulations of galaxies to the velocity maps from Gaia in order to get insights on the bar-spiral orientation. Our approach hereafter is complementary since, instead of qualitatively comparing a self-consistent hydrodynamical simulation to the data, we attempt a more quantitative fit to the stellar phase-space data from Gaia. For this, we will resort to backward integrations to model the velocity field with a parametric form of the gravitational potential. Our preferred solution could then serve as new fiducial non-axisymmetric parametric potential for the Milky Way disk. 

In Section~\ref{sec:data}, we briefly present the data extracted from the RVS sample of Gaia DR3 that we are using to constrain the potential from the MW disk kinematics. The modeling method and the parametrization of the potential are introduced in Section~\ref{sec:backwards_integration}, whilst results are presented in Section~\ref{sec:results}. {\it A posteriori} comparisons with observables to which the model was not fit, as well as some examples of applications of our fiducial potential are given in Section~\ref{sec:appli}, before concluding in Section~\ref{sec:concl}.

%--------------------------------------------------------------------
\section{The Gaia RVS disk sample}
\label{sec:data}
\begin{figure*}[h]
\centering
\includegraphics[width=\hsize]{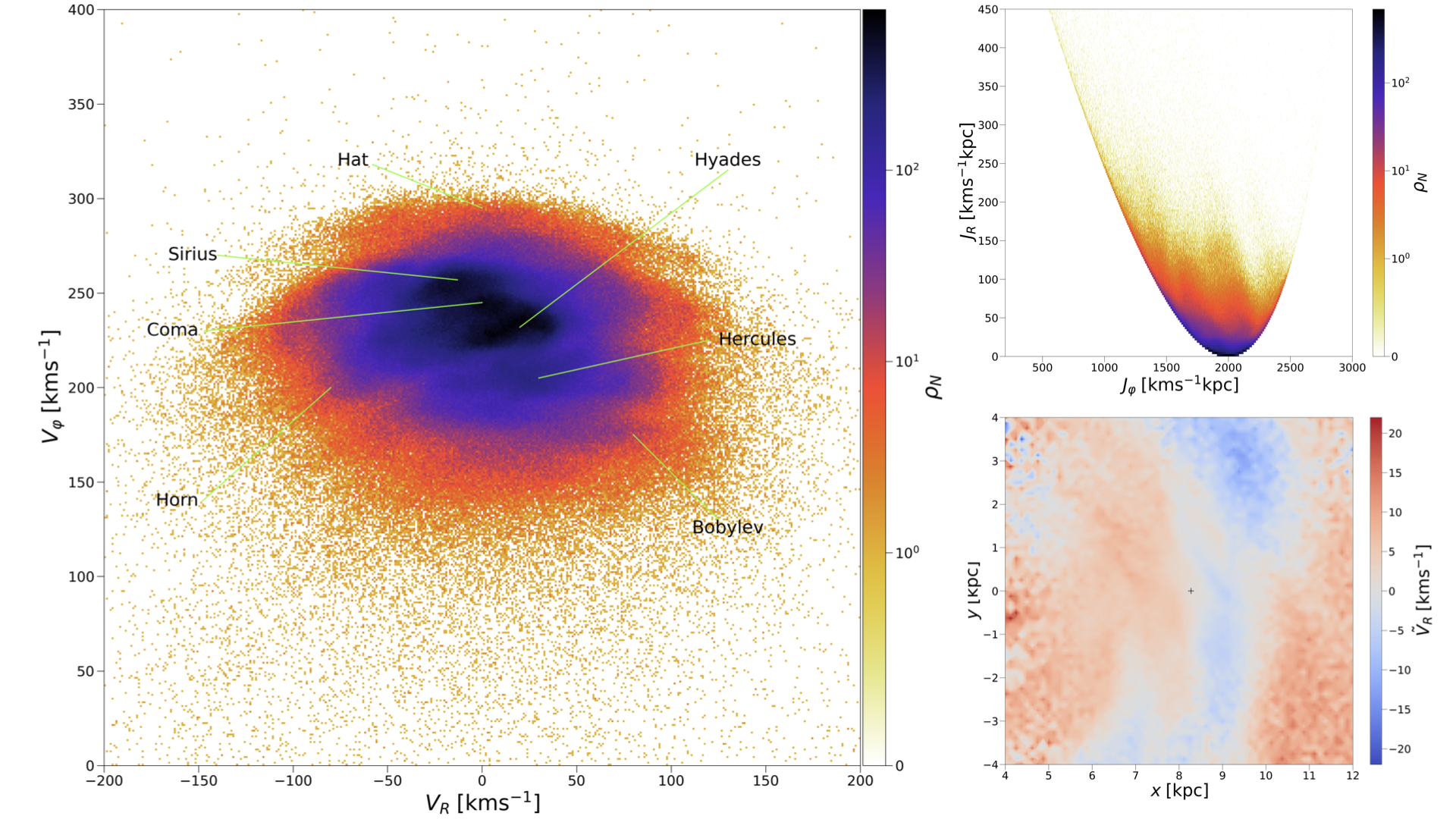}
\caption{
Left panel: 2-dimensional histogram of the number density of stars from the Gaia RVS disk sample (in a cylinder of 300~pc radius and $\pm 300 \, \pc$ height around the Sun) in the $(V_{R},V_{\varphi})$ plane defined on a $[-200,200]~\kms\times[0,400]~\kms$ grid, binned with $(1~\kms)^2$ bins, together with the locations of the different moving groups. 
Top right panel: Same distribution in the $(J_{\varphi},J_{R})$ plane (computed with {\tt AGAMA} in the potential of Table~\ref{table:axisymmetric}) defined on a $[0,600]~\kms \, {\rm kpc}^{-1}\times[200,3000]~\kms \, {\rm kpc}^{-1}$ grid,  binned with $(3 \times 2 \; {\rm km^2 \, s}^{-2} \, {\rm kpc}^{-2})$ bins. 
Bottom right panel: the median radial velocity $\tilde{V}_{R}$ $(x,y)$ as a function of the position in the Galactic plane for the full sample of 17~414~667 stars with $|z| < 300$~pc. The grid is defined as $[4, 12]~{\rm kpc} \times [-4, 4]~{\rm kpc}$, binned with $(125 \pc)^2$ bins. 
The Galactic center is located at $(x,y)=(0 \, {\rm kpc}, 0 \, {\rm kpc})$, the Sun at $(x,y)=(8.275 \, {\rm kpc}, 0 \, {\rm kpc})$ is represented with a cross, and the sense of rotation of the Galaxy is anti-clockwise. 
}
\label{fig:data}
\end{figure*}

Since we are planning to use the in-plane motions of disk stars to constrain the non-axisymmetries of the MW, we select a sample of stars with six-dimensional phase-space information from the Gaia RVS close to the Galactic plane. We use data from Gaia DR3 \citep{Gaia2023a} combined with the StarHorse \citep{Anders2022} distances, and select 17 414 667 stars within a height of 300~pc from the Galactic plane. 

We adopt, for the Sun's position $\bx_{\odot}$ and velocity $\bv_{\odot}$ in Galactocentric Cartesian coordinates, $\bx_{\odot}=(x_\odot,y_\odot,z_\odot)=(8275, 0, 15.29)\,\pc$ and $\bv_{\odot} = (V_{x_\odot},V_{y_\odot},V_{\rm z_\odot})=(-9.3, 251.5, 8.59) \, {\rm km}\, {\rm s}^{-1}$ \citep{Gaia2023b,Portail2017,Widmark2019}, respectively. We then transform the data from equatorial coordinates to Galactocentric coordinates with the Astropy library \citep{Astropy2022} to compute the stars' positions in Galactocentric Cartesian coordinates, $\bx=(x,y,z)$ and their in-plane velocities in Galactocentric Cylindrical coordinates, $\bv=(V_R,V_{\varphi})$, with the Galactocentric radius $R = \sqrt{x^2 + y^2}$ and azimuth $\varphi = {\rm arctan}(y/x)$, defined to be zero at the azimuth of the Sun and positive in the direction of Galactic rotation. The stars are selected within $4 \, {\rm kpc} < x < 12 \, {\rm kpc}$ and $-4 \, {\rm kpc} < y < 4 \, {\rm kpc}$. 

On Fig.~\ref{fig:data}, left panel, we display the local stellar velocity distribution for a bit more than 2 million stars within a cylinder of radius $300 \, \pc$ around the Sun, still within a height of $300 \, \pc$. In this figure several of the well-known moving groups of the Solar neighborhood \citep[e.g.,][]{Dehnen1998,Famaey2005,Antoja2008,Ramos2018,Bernet2022} are immediately visible. The ``Hat'' can be seen as the downward concave arch at high $V_\varphi$, from $(V_{R}, V_{\varphi}) \approx (-100~\kms, 270 ~\kms)$ to $(V_{R}, V_{\varphi}) \approx (120 ~\kms, 260 ~\kms)$; The Sirius moving group \citep[e.g.,][]{Famaey2008} is approximately straight at $V_{\varphi} \approx 255 ~\kms$, located between $V_{R} \approx -50 ~\kms$ and $V_{R} \approx 0 ~\kms$, with a peak at $V_{R} \approx -15 ~\kms$; Coma is right below Sirius in azimuthal velocity, around $(V_{R}, V_{\varphi}) \approx (0 ~\kms, 245 ~\kms)$; The Hyades moving group \citep[e.g.,][]{Famaey2007,Pompeia2011} can be seen as a slightly curved downward arch from the over-density at $(V_{R}, V_{\varphi}) \approx (20 ~\kms, 230 ~\kms)$; The Horn is right next to the Hyades, on the other side in $V_R$: it appears as an arch going through $(V_{R}, V_{\varphi}) \approx (-80 ~\kms, 200 ~\kms)$; finally, the major Hercules moving group is perceived as a bimodality of the whole velocity-plane, with an under-density, just below the Hyades in azimuthal velocity, separating it from the rest of the distribution. Its bimodality appears clearly, with a second overdensity appearing at low $V_\varphi$.

Another way to visualize these arches, which however visually erases the asymmetries in radial velocity, is to plot the distribution of stars in the local axisymmetric action space \citep[e.g.][]{Trick2019,Trick2021}. Indeed, the Galaxy is to leading order an axisymmetric quasi-integrable system, and the action-angle variables $(\bJ, \btheta)$ are the canonical phase-space variables appropriate for studying and perturbing integrable systems. In these coordinates, the Hamiltonian only depends on the actions $\bJ$, which are integrals of the motion. Each triplet of actions then fully characterizes an integrable orbit, while the angles indicate where a given star is on that particular orbit. The azimuthal action is simply $J_\varphi=R \, V_\varphi$, whilst the radial action $J_R$ (computed within the background axisymmetric potential defined in Sect.~3) encodes the (Galactocentric) radial excursions of a given orbit. In the top-right panel of Fig.~\ref{fig:data}, the arches in local velocity space are now seen as ridges in local action space, characteristic of resonant features \citep[e.g.,][]{Monari2017a,Binney2020}.

These features of local velocity and action space, traced with exquisite precision, have however been known for a long time. The most interesting added value of Gaia data releases has been to expand the volume around the Sun where such dynamical features can be studied \citep[e.g.,][]{Ramos2018,Bernet2022}. In order to adjust the non-axisymmetric components of the Galactic potential in the present work, we will however refrain from using the full phase-space distribution of disk stars, and will rather fit the measure of a central tendency as a function of position in the disk, namely the median Galactocentric radial velocity \citep{Gaia2023b}. This map of median radial velocity is displayed in the bottom-right panel of Fig.~\ref{fig:data} and will be the main observable adjusted in the present work. We will check only {\it a posteriori} the qualitative agreement with the full phase-space distribution of stars. Since our modeling will be based on a projected 4D phase-space distribution function (DF, see Sect.~3) of the disk stellar populations -- a DF that is supposed to take into account stars with non-zero vertical velocities --, we do not make any additional cuts on the vertical velocity in the data. However, while our DF is a projected one, our orbit integrations will be performed only within the plane. Therefore, we also checked that selecting only stars with vertical velocities below $15 \, {\rm km}\, {\rm s}^{-1}$, allowing to keep a reasonable number of $11~427~688$ stars in the dataset, led to an almost identical median radial velocity map. For the important points of the fit, the typical differences are below 0.5 km/s, with a maximum difference of 1 km/s.

%-------------------------------------- 
\section{Modeling} \label{sec:backwards_integration}
\begin{figure*}[h]
\centering
\includegraphics[width=\hsize]{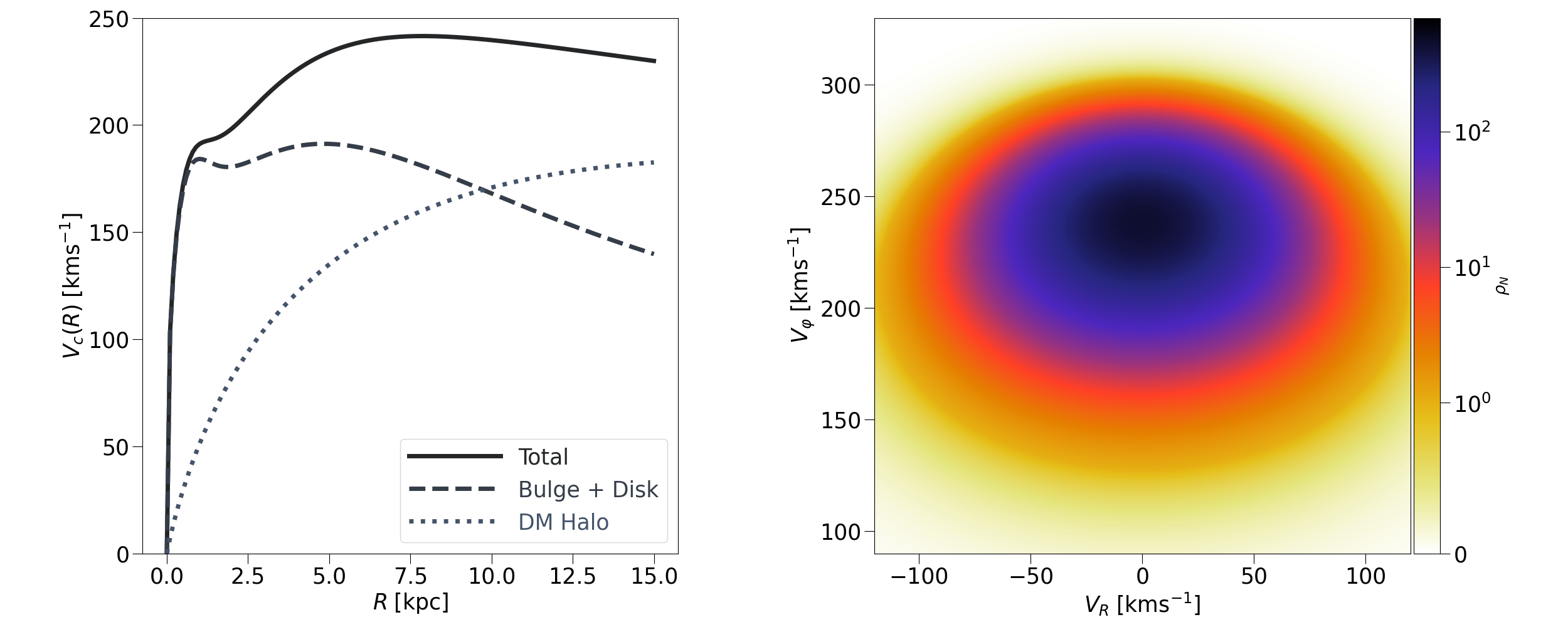}
\caption{
The axisymmetric background. Left panel: the circular velocity curve of the background axisymmetric potential described in Sec.~\ref{sec:background_potential}. 
Right panel: The number density $\rho_N$ of stars in velocity space at the Sun's position from the equilibrium DF described in Sec.~\ref{sec:diskDF}, for a normalization factor such that the total number in the model at the Sun is the same as found in the data within the 300~pc cylinder around the Sun.
}
\label{fig:background}
\end{figure*}

To build our non-axisymmetric potential we start from an axisymmetric one, and subsequently add a bar and spiral arms, defined by several parameters as described in the following subsections. Within the axisymmetric potential, we use a phase-space DF, $f(\bx,\bv)$, to describe our tracer population. This 1-particle DF is the probability density function of finding one star at the phase-space point $(\bx,\bv)$ and, for a collisionless system, it obeys the collisionless Boltzmann (or Vlasov) equation. Such a DF, for any steady-state integrable stellar system, should solely be a function of isolating integrals of the motion according to the Jeans theorem \citep{Binney2008}. We take these integrals to be the actions $\bJ$. In order to transform stellar positions and velocities into actions, one can use an approximation based on Stäckel potentials \citep[see, e.g.,][]{Famaey2003}. Axisymmetric Stäckel potentials are expressed in a spheroidal coordinate system, defined by a focal distance, which is always related to the first and second derivatives of the potential at any given position. Thus, by using the actual Galactic potential at any given position, one can compute an equivalent focal distance as if the potential were locally of Stäckel form, allowing for the calculation of the (quasi-)integrals of the motion and corresponding actions. This ``Stäckel fudge" \citep{Binney2012,Sanders2016} is fully implemented within the the Action-based Galaxy Modeling Architecture code \citep[][{\tt AGAMA}]{Vasiliev2018,Vasiliev2019} that we use in the present study.

We are starting from an equilibrium DF, $f(\bJ)$, for the tracer population defined within an axisymmetric potential. The first route to include the effect of non-axisymmetric components is to treat them through perturbation theory \citep[e.g.][]{Monari2016a,Kazwini2022} which, in order to be truly quantitative, needs a special treatment for the resonant zones for a constant pattern speed perturbation \citep[e.g.][]{Monari2017a,Laporte2020,Binney2020,Hamilton2023}. This becomes however practically very complicated in the presence of multiple perturbers with different pattern speeds, whose resonances will overlap. To circumvent this issue, one can fortunately rely on the property encoded in the collisionless Boltzmann equation, that the DF value in an infinitesimal Lagrangian volume is conserved along the trajectory. This allows us to use the classical method of backward integration \citep{Vauterin&Dejonghe1997}, used in, e.g., \citet{Dehnen2000,Hunt2018a,Hunt2018b,Hunt2019,Monari2019a,Bernet2024}, in order to explore the shape of the DF.

In order to compute the $f(\bx,\bv,t)$, at current time $t=0$, at the phase-space point $(\bx,\bv)$, in the presence of the bar and spiral arms, we backward integrate the orbit for a fixed integration time to its phase-space position $(\bx',\bv')$ at time $t'<0$, before the actual appearance of the non-axisymmetric perturbers. Assuming that the tracer population is represented by the equilibrium DF, $f(\bJ)$, in the axisymmetric background potential at time $t'$, we transform $(\bx',\bv')$ to action-angle variables using {\tt AGAMA}, compute the value of the DF\footnote{Let ${\cal J}$ be the Jacobian of the transformation between $(\bJ,\btheta)$ and $(\bx,\bv)$. Since $\text{det}{\cal J}=1$ for canonical variables \citep{Binney2008}, the transformation between a DF in $(\bJ,\btheta)$, $f(\bJ,\btheta)$, and the one in $(\bx,\bv)$, $f'(\bx,\bv)$, is given by $f'(\bx,\bv)=f(\bJ(\bx,\bv),\btheta(\bx,\bv))~\text{det}{\cal J}=f(\bJ(\bx,\bv),\btheta(\bx,\bv))$, i.e. $f'$ is just $f$ with $\bJ$ and $\btheta$ written as functions of $\bx$ and $\bv$.}, and since this value in an infinitesimal Lagrangian volume is conserved, we attribute the same value of the DF to the phase-space position $(\bx,\bv)$ at present time $t=0$ in the presence of the bar and spirals. In practice, the orbits are integrated within the plane only, by solving the initial value problem with the Runge-Kutta of order 5 method {\tt odeint} solver from the very efficient {\tt torchdiffeq} library \citep{torchdiffeq} in {\tt PyTorch} \citep{PyTorch}. Doing this at numerous phase-space locations allows us to compute the median radial velocity as a function of position in the disk, and to adjust the parameters of the non-axisymmetric components in order to fit the observed values. In practice the median radial velocity at each grid position (sampled every 50~pc in $x$ and $y$) on the disk is computed after locally integrating the values of the DF in $V_{\varphi}$ for a grid of velocities on which the backward integration is performed at each position. This grid ranges from $-79 \kms$ to $79 \kms$ with a step of $2 \kms$ in $V_R$, and from $110 \kms$ to $330 \kms$ with a step of $4 \kms$ in $V_\varphi$. The potential is evaluated on a grid of radii that is subsequently interpolated with a cubic spline in the {\tt torchcubicspline} library in {\tt Pytorch} to improve computational time. Similarly, we also interpolate, with {\tt Scipy} \citep{Scipy}, a cubic spline to the actions computed with {\tt AGAMA}.

There are three caveats with the method, which are worth mentioning even though addressing them in detail is far beyond the scope of this first quantitative approach to the problem. First, we are using the full disk sample described in the previous section without taking into account a detailed selection function, assuming that the high number of stars that we use allows for a good estimate of the true median velocity. The second caveat is that, in practice, the observed stellar DF is always measured over finite phase-space volumes whilst the backward integration method operates under the assumption that the mean value of the DF within a given phase-space volume is equivalent to its value at the central point, irrespective of how the volume deforms during the system's orbital evolution. In other words, the backward integration method yields the fine-grained DF, which will typically remain unsmoothed at small scales, whilst the measurable DF within observations is the coarse-grained one, which does {\it not} obey the collisionless Boltzmann equation (as this coarse-grained DF is smoothed by phase-mixing within finite volumes). The Nyquist-Shannon sampling theorem imposes limits on the minimum size of fine structures in phase space that can form for a fixed number of particles over time, and this limit is reached on rather short time scales, shorter than collisional relaxation \citep{Beraldo}. Once this limit is reached, the system cannot form finer structures, despite the collisionless Boltzmann equation predicting that these structures do form. In practice, this means that, if the integration is carried out for too long, the fine-grained DF tracked by the backward integration method will lead to sharp and unsmoothed features in velocity space, where chaotic features will also appear as sharper than in the real world. To circumvent this problem, the integration must be carried out only for a relatively limited time, adjusted so that the sharpness of resonant features in velocity space resembles what is observed. Luckily, $N$-body simulations indicating the existence of recurrent cycles of groove modes in galactic disks \citep{Sellwood2014,Sellwood2019} allow us to consider that current spiral arm modes of the Milky Way are rather recent. This assumption is, of course, not ideal for the bar, but it is reasonable to assume that the {\it location} of the resonant features in local velocity space will not evolve with time, whilst their sharpness will. Hence, we will only deal with the location of resonant features in local velocity space to constrain the pattern speed of the bar, and rely on a parametric form of its potential adjusted to the dynamics of the bulge region \citep{Portail2017,Thomas23} for its amplitude. It would be too costly to resort to a forward integration method within the fitting scheme that we set out to apply in the present paper, given the size of the parameter space to explore, and given that each combination of parameters requires a full backward integration of the whole Galactic plane. However, the results obtained hereafter could serve as a basis for forward-in-time test-particle simulations, also expanded to three-dimensions, that we plan to present in a follow-up paper. Finally, a third and last caveat is that our simulations are, by design, not self-consistent. This simplification is much more efficient to explore a vast parameter space. However, future improvements of our method might rely on an adaptation of the made-to-measure method \citep{Syer96,Portail2017} to account for self-consistency, using the results presented hereafter as a basis. 

\subsection{Background axisymmetric potential} \label{sec:background_potential}
As outlined hereinabove, the method uses an axisymmetric background potential. In practice, we assume a 3D axisymmetric density profile and the potential is computed by solving Poisson's equation with {\tt AGAMA}. The density profile is the summed density of each of the following components: stellar disk, gas disk, bulge and dark matter halo. 

The stellar and gas disk density profiles are parametrized in Galactocentric cylindrical coordinates $(R,z)$ as: 
\begin{align}\label{eq:rho_disk}
\rho_{\rm disky}(R,z) = \frac{\Sigma_0}{2h_z} {\rm exp}\left( - \Bigl|\frac{z}{h_z}\Bigr| \right) {\rm exp} \left( - \frac{R}{h_R}\right),
\end{align}
with the central surface density $\Sigma_0$, scale height $h_z$ (and hence central density $\Sigma_0/2h_z)$, and scale length $h_R$. The spherical density profile for the bulge and dark matter halo is given by: 
\begin{align}\label{eq:rho_spherical}
\rho_{\rm spheroidal}(R,z) = \rho_0  \left( \frac{\tilde{r}}{a} \right)^{-\gamma}
\left( 1 + \frac{\tilde{r}}{a} \right)^{\gamma - \beta}
{\rm exp} \left[ - \left( \frac{\tilde{r}}{r_s} \right) ^{\alpha} \right],
\end{align}
with a density normalization $\rho_0$, a scale radius $a$, an outer scale radius $r_s$, and exponents $\alpha$, $\beta$, $\gamma$. The ellipsoidal radius is defined as $\tilde{r} = \sqrt{R^2 + \left( \frac{z}{q} \right)^2}$, with $q$ the vertical axis ratio. All parameters are given in Table~\ref{table:axisymmetric}. The baryonic mass of the model is $6 \times 10^{10} {\rm M}_\odot$ and the dark matter halo is relatively light, with a mass of $3.1 \times 10^{11} {\rm M}_\odot$, in between the typical values obtained from circular velocity curve analyses \citep[e.g.,][]{Jiao,Ou} and those obtained from escape speed curves, satellite dynamics or stream fitting \citep[e.g.,][]{Monari18,Callingham,Roche,Ibata24}. Only the mass in the inner Galaxy however matters for our present modeling: the total enclosed mass (baryons and dark matter) within 20~kpc is $2.2 \times 10^{11} {\rm M}_\odot$, roughly in agreement with the \citet{Malhan19} constraint. The local dark matter density at the Sun's position is $1.3 \times 10^{-2} {\rm M}_\odot \pc^{-3}$, consistent with most estimates \citep[][and references therein]{deSalas}. In the center, the dark matter halo displays a constant density core (with a central power-law slope of 0) as well as a shallow power-law decline close to the center with a slope of -0.6 at $R=1$~kpc and of -1 at $R=3$~kpc. All these background potential parameters could in principle be let free in our fitting procedure hereafter, but to simplify the problem they have all been fixed to resemble closely the axisymmetric part of the model by \citet{Portail2017}. The circular velocity curve corresponding to this axisymmetric model  is plotted in the left panel of Fig.~\ref{fig:background}. The non-axisymmetric modes that will be added on top of this axisymmetric background will all have zero total mass, meaning that the total mass of the final non-axisymmetric model will be the same as that of the axisymmetric one. Since our orbits will be computed strictly within the plane, we only need hereafter the background potential within the plane, $\Phi_0(R)$.

\begin{table*}[h]
\caption[]{\label{table:axisymmetric} Fixed parameters of the axisymmetric background density used to compute the background potential, following the definitions in Eq.~\ref{eq:rho_disk} and Eq.~\ref{eq:rho_spherical}.}
\centering
\begin{tabular}{lcccccccc}
\hline \hline
    & & & \textbf{Disky density profiles} & & & & \\ \hline
    Component & $\Sigma_0$ ($10^3 {\rm M}_\odot \, \pc^{-2}$)& $h_R$ (kpc) & $h_z$ (kpc) & & & & \\ \hline
    Stellar disk & $1.19$ & $2.4$ & $0.30$ & & & & \\
    Gas disk & $0.07$ & $4.8$ & $0.13$ & & & & \\
\hline \hline \\
    & & & \textbf{Spheroidal density profiles} & & & &  \\ \hline
    Component & $\rho_0$ ($10^{-1} {\rm M}_\odot \, \pc^{-3}$)& $a$ (kpc) & $r_s$ (kpc) & $\alpha$ & $\beta$ & $\gamma$ & $q$  \\ \hline
    Bulge & $1.08 $ & $8.16$ & $0.83$ & $2.0$ & $2.9$ & $1.3$ & $1.0$ \\ 
    DM halo &  $4.56 $ & $-$ & $0.65$ & $0.50$ & $0.0$ & $0.0$ & $0.8$ \\
\hline
\end{tabular}
\end{table*}

\subsection{Axisymmetric equilibrium distribution function}
\label{sec:diskDF}

The second step of our procedure consists in choosing an equilibrium DF for the tracer stellar population within the plane. Since we are confined to the plane, we do not attempt here to be fully self-consistent \citep[see, e.g.][]{Binney2023}, in order to allow for a simple and tractable form of the DF, namely a simple linear combination of two quasi-isothermal DFs $f(J_R,J_\varphi) = F_{\rm thin} + \zeta F_{\rm thick}$ with $\zeta = 0.05$, that are two-dimensional in action space, and both with the following form \citep{Binney2010,Binney2011},
\begin{align}\label{eq:DF}
F = \eta \frac{\Omega(J_\varphi)}{\kappa(J_\varphi) \tilde{\sigma}^2_R(J_\varphi)} {\rm exp} \left( -\frac{R_\mathrm{g}(J_\varphi)}{h_R} \right) {\rm exp}\left( - \frac{J_R\kappa(J_\varphi)}{\tilde{\sigma}^2_R(J_\varphi)} \right),
\end{align}
with $R_\mathrm{g}$ the guiding radius, $\Omega, \kappa$ the circular and epicyclic frequencies, all three depending on the azimuthal action $J_\varphi$, $h_R$ the disk scale length, $\eta$ the normalization factor (in units of inverse length squared) of the tracer population, and finally the radial velocity dispersion $\tilde{\sigma}_R$ depending on the guiding radius as:
\begin{align}\label{eq:sigma_R}
\tilde{\sigma}_R(R_\mathrm{g}(J_\varphi)) = \tilde{\sigma}_R(R_0) {\rm exp} \left( - \frac{R_\mathrm{g}(J_\phi) - R_0}{h_{\sigma, R}} \right),
\end{align}
where $h_{\sigma, R}$ is the kinematic scale-length of the tracer population. For  $F_{\rm thin}$, we set the scale length to $h_{R} = 2.4$~kpc in accordance with the potential, the velocity dispersion at the Sun's position to $\tilde{\sigma}_{R,{\rm thin}}(R_0) = 30$~kms$^{-1}$, and the kinematic scale length to $h_{\sigma_R} = 10$~kpc. For $F_{\rm thick}$, the only difference is that we set $\tilde{\sigma}_{R,{\rm thick}}(R_0) = 55$~kms$^{-1}$. Our DF corresponds to a projected four-dimensional DF in phase-space, namely in units of inverse length-squared times inverse velocity-squared, hence corresponding to the 6D DF of the modeled disk populations integrated over heights and vertical velocities. The local velocity distribution at $R=R_0$ corresponding to this axisymmetric DF is displayed in the right panel of Fig.~\ref{fig:background}. In practice, the normalization factor is adjusted such that the number of stars in the model at the Sun is the same as found in the data within the cylinder of 300~pc radius and $\pm 300 \, \pc$ height around the Sun.

\subsection{Non-axisymmetric potential} \label{non_axisymmetric_potential}

The third step of our procedure is to add non-axisymmetric modes on top of the axisymmetric background potential $\Phi_0$. The total potential is obtained by adding to $\Phi_0(R)$ the real part of the following:
\begin{align}\label{eq:pot_tot}
\Phi_\text{tot}(R,\varphi,t)&=  \Phi_0(R) +\sum_{m} \phi_{{\rm b},m} (R,t) \, {\rm exp}[{\rm i} \, m(\varphi - \varphi_\text{b,0} -\Omega_\text{b}t)] \nonumber \\
&\quad +\sum_{m} \phi_{{\rm s},m} (R,t) \, {\rm exp}[{\rm i} \, m(\varphi - \varphi_{{\rm s},m,0} -\Omega_{{\rm s},m}t)],
\end{align}
 where the current phase and the pattern speed of the bar are respectively $\varphi_\text{b,0}$ and $\Omega_\text{b}$, and those of the spiral arms mode $m$ respectively $\varphi_{{\rm s},m,0}$ (the present-day spiral phase at the Solar position) and $\Omega_{{\rm s},m}$. 
 The amplitude of each mode is given by $\phi_{{\rm b},m}$ and $\phi_{{\rm s},m}$ for the bar and spirals, respectively. The time $t$ is such that currently $t = 0$.

As outlined hereinabove, the amplitude of the modes of the bar potential are fixed to values that fit well the dynamics of the bulge region. Namely, the bar potential is a superposition of three Fourier modes, with the same parametric form as in \citet{Thomas23}, closely resembling the three first even modes of the bar potential derived in \cite{Portail2017}. From this same potential, the bar angle phase is fixed to be $\varphi_\text{b,0} = 28^{\circ}$. The amplitude of each bar mode $m$ is given by:
\begin{equation}
\phi_{{\rm b},m} (R,t)= G_{\rm b}(t) \, A_{{\rm b},m} (R) \, \Phi_0(R),
\end{equation}
where $G_{\rm b}(t) \leq 1$ is the growth function for the bar, and $A_{{\rm b},m}$ is the relative amplitude of the bar mode given by
\begin{equation}\label{eq:bar_amplitude}
A_{{\rm b},m} (R)= K_{{\rm b},m}~(R/R_\text{b,max})^{a_m-1}(1-R/R_\text{b,max})^{b_m-1},
\end{equation}
with $K_{{\rm b},m}$ a global amplitude factor and $R_\text{b,max}$ the radius at which the mode's amplitude goes to zero. Importantly, we consider that the amplitude has reached a plateau at the present time $G_{\rm b}(t=0)=1$. The values of $K_{{\rm b},m}, ~a_m,$ and $b_m$ for each of the bar modes are presented in Table~\ref{table:non-axisymmetric}. Only the pattern speed of the bar is adjusted to the location of resonant ridges in local velocity space within our procedure (see next Section).

The spiral arms potential that we propose is an adaption of the analytical model of \cite{Cox2002} described in \cite{Monari2016b}, whose amplitude is given by 
\begin{equation}\label{eq:spiral_arms}
\phi_{{\rm s},m} (R,t)= G_{\rm s, m}(t) \, A_{{\rm s},m} (R) \, {\rm exp}\left[{\rm i} \, m \frac{{\rm ln}(R/R_0)}{{\rm tan} p_{\mathrm{s},m}}\right] \, \Phi_0(R),
\end{equation}
where $G_{\rm s, m}(t)$ is the growth function for the spiral arms mode $m$, set to $G_{\rm s, m}(t=0)=1$, $p_{\mathrm{s},m}$ is the pitch angle, and $A_{{\rm s},m}$ is given by
\begin{equation}\label{eq:arms_amplitude}
A_{{\rm s},m} (R)= \xi_{{\rm s},m}(R) \, H_{m}(R) \, \frac{\Phi_0(R_0)}{\Phi_0(R)},
\end{equation}
where $\xi_{{\rm s},m}$ is the amplitude factor of the mode, normalized to its value $K_{{\rm s},m}$ at $R = R_0$ with a radial dependence as follows:
\begin{equation}
\begin{split}
\xi_{{\rm s},m}(R) = K_{{\rm s},m} \frac{R_0^2 \sin^2 p_{s,m} + m h_{\mathrm{s},m} R_0 \sin p_{s,m} + 0.3 m^2 h_{\mathrm{s},m}^2}{ R^2 \sin^2 p_{s,m} + m h_{\mathrm{s},m} R \sin p_{s,m} + 0.3 m^2 h_{\mathrm{s},m}^2}\\ \times \frac{R^3 \sin p_{s,m} + 0.3 m h_{\mathrm{s},m} R^2}{R_0^3 \sin p_{s,m} + 0.3 m h_{\mathrm{s},m} R_0^2}.
\end{split}
\end{equation}
This adaptation of the \cite{Cox2002} potential has the advantage to be easily generalizable to 3D. Here, $h_{\mathrm{s},m}$ corresponds the scale-height of the spiral potential, which we fix to 130~pc. We have checked that our results are not very sensitive to this parameter and are similar for any values between 100~pc and 300~pc. Finally, $H_{m}$ is a radial cutoff function, parametrized by an inner and an outer cutoff, respectively, $R_{{\rm s},m,{\rm min}}$ and $R_{{\rm s},m,{\rm max}}$. The function is simply\footnote{This cut-off is chosen for simplicity, in a context where we do not aim for self-consistency. However, in a context where the density-potential pair of a spiral mode is needed at the cut-off, it is desirable to replace the step function by something like $H_m \sim 0.5(1+{\rm tanh}((R-R_{{\rm s},m,{\rm min}})/\Delta_{\rm cutoff}))$, where $\Delta_{\rm cutoff} \rightarrow 0$ corresponds to our present case.}:
\begin{equation}
    H_{m}(R) = \begin{cases}
  1 & \text{if } R_{{\rm s},m,{\rm min}} \leq R \leq R_{{\rm s},m,{\rm max}}, \\
  0 & \text{otherwise}.
\end{cases}
\label{step}
\end{equation}
These cutoffs will be determined from the pattern speeds of the bar and spiral modes in the next Section. The parameters of the spiral arms (for each mode: amplitude $K_{s,m}$, pitch angle $p_{s,m}$, present-day phase at the Solar position $\varphi_{{\rm s},m,0}$, and pattern speed $\Omega_{{\rm s},m}$) will be adjusted to the data in the next Section, together with the bar pattern speed $\Omega_\text{b}$.
%-------------------------------------- 
\section{Fitting procedure and results} \label{sec:results}
\subsection{Bar-only model} \label{bar_only}

\begin{figure}
\centering
\includegraphics[width=\hsize]{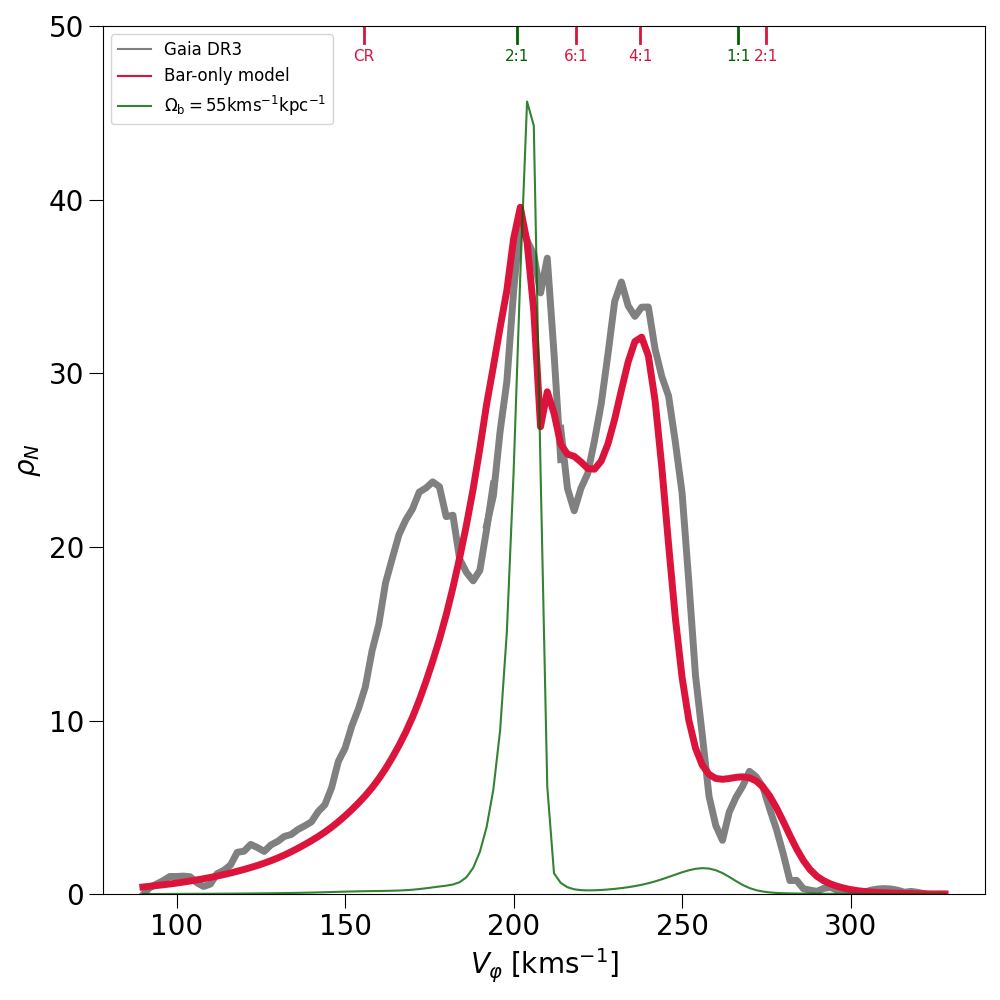}
\caption{
1-D distribution of the azimuthal velocity of stars in the Solar neighborhood within $99 ~\kms< V_R < 101 ~\kms$, a region of velocity space where the bar impact dominates the distribution (over potential spiral arms signatures). In grey, the stellar distribution from the Gaia RVS disk sample in the Solar neighborhood, smoothed with the Savitzky-Golay filter from the {\tt SciPy} library. Red line: the renormalized best bar-only model at $V_R = 100 \, \kms$, with pattern speed  $\Omega_{{\rm b}} = 37 \, \kmskpc$. For reference, we are providing the results for $\Omega_{{\rm b}} = 55 \, \kmskpc$  (green line), where only the 1:1 resonance leaves a small signature at higher $V_\varphi$ than the strong OLR peak. The approximate locations of the different resonances evaluated from Eq.~\ref{eq:dehnen_vphi_resonant} are represented as red ($\Omega_{{\rm b}} = 37 \, \kmskpc$) and green ($\Omega_{{\rm b}} = 55 \, \kmskpc$) dashes on top of the plot.
}
\label{fig:bar_peaks}
\end{figure}
\begin{figure*}
\centering
\includegraphics[width=18cm]{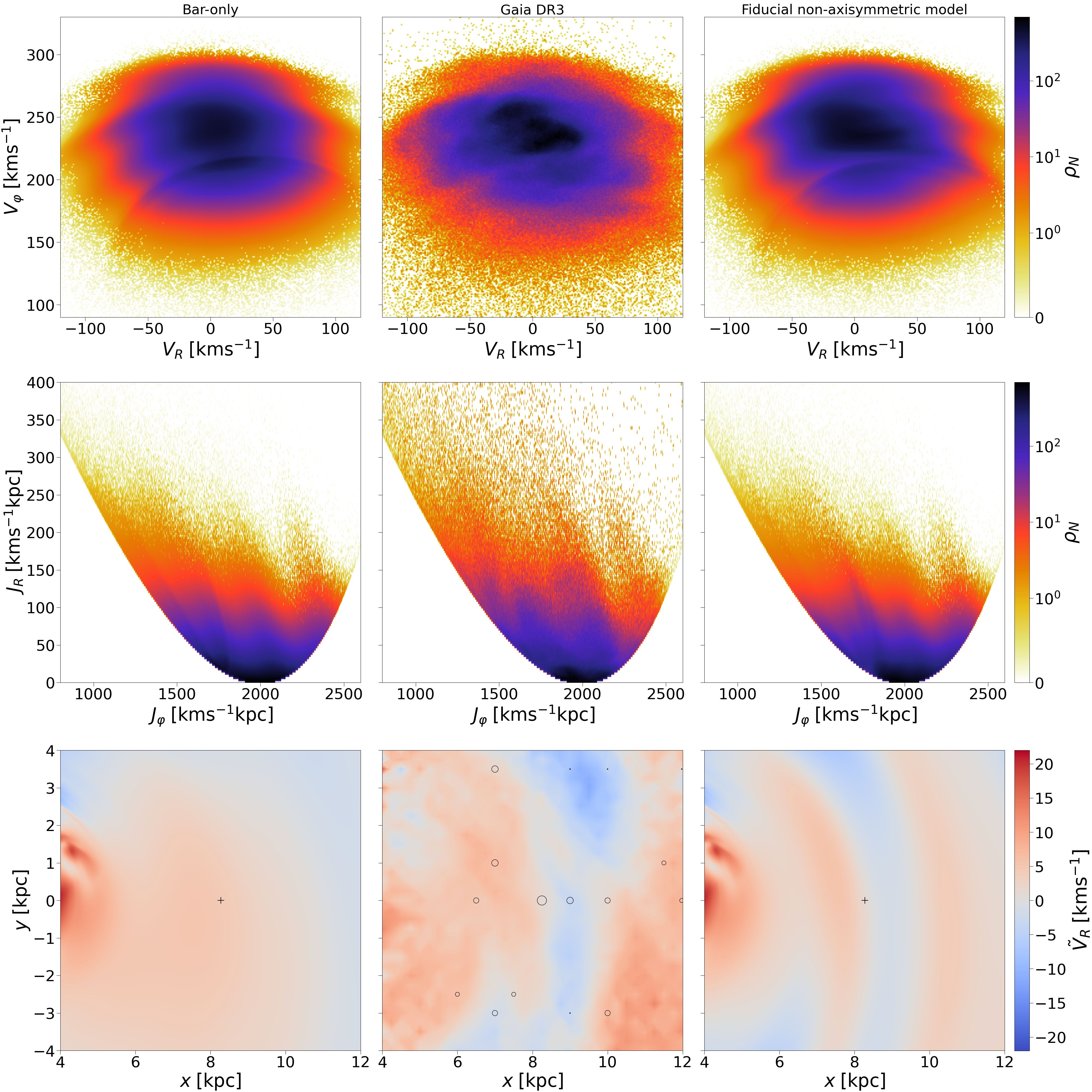}
\caption{
From left to right, the columns correspond to the bar-only model, the Gaia RVS disk sample and our fiducial non-axisymmetric model, respectively.
Top row: the  2-dimensional histogram of stars in the local $(V_{R},V_{\varphi})$ plane defined on $[-200,200]\kms\times[0,400]\kms$, binned with bins of size $(1 \kms)^2$. Middle row: the  2-dimensional histogram of the number density of stars in the $(J_{R},J_{\varphi})$ plane defined on $[0,400]\kmskpc\times[800,2600]\kmskpc$, binned with bins of size $(3 \times 2 \; {\rm km^2 \, s}^{-2}{\rm kpc}^{-2})$. For this, the velocities $(V_{R},V_{\varphi})$ have been transformed to actions $(J_{R},J_{\varphi})$ with {\tt AGAMA}. Bottom row: the median $\tilde{V}_{R}$ $(x,y)$ is shown in the $(x,y)$ plane defined on $[4, 12]\kpc \times [-4, 4]\kpc$, and binned with bins of size $(250 \pc)^2$ for the data and $(50 \pc)^2$ for the models. The cross locates the Sun and circles (with sizes proportional to the weights, the lowest weights being dots) in the middle panel (data) indicate the selected points where the fit has been performed. All panels were smoothed with a bi-linear interpolation. } 
\label{fig:results}
\end{figure*}

\begin{table*}[h]
 \caption[]{\label{table:non-axisymmetric}Parameters of the planar fiducial non-axisymmetric potential fitted in Sect.~4.2, with parameters as defined in Eqs \ref{eq:bar_amplitude} and \ref{eq:arms_amplitude}. The surface density contrasts at the Solar radius, $\Sigma_{\rm s}$, are computed from the integrated 3D density equation for spiral arms in \citet{Cox2002} over the baryonic surface density of the background.}
 \centering
\begin{tabular}{ccccccccccccccc} 
 \hline \hline
 & & & & & \textbf{Bar} & & & & & & \\
 \hline
  $\Omega_\text{b}$ &
  $\varphi_\text{b,0}$ &
 $R_\text{b,max}$ &
 $K_{{\rm b},2}$ &
 $K_{{\rm b},4}$ &
 $K_{{\rm b},6}$ &
 $a_{2}$ &
 $a_{4}$ &
 $a_{6}$ &
 $b_{2}$ &
 $b_{4}$ &
 $b_{6}$ \\
\hline
$37\kmskpc$ & 28$^\circ$ & $12 \kpc$ & 0.25 & 8.4 & 210.41 & 1.8 & 4.08 & 5.96 & 5.08 & 10.7 & 16.06 \\ 
\hline
\hline
\\
 & & & & & \textbf{$\mathbf{\bsym{m}}$=2 spiral} & & & & & & & \\
 \hline
  $\Omega_{{\rm s},2}$ &
  $K_{{\rm s},2}$ &
  $\varphi_{{\rm s},2}$ &
  $p_{{\rm s},2}$ &
  $R_{{\rm s},2,{\rm min}}$ &
  $R_{{\rm s},2,{\rm max}}$ &
  $h_{{\rm s},2}$&
  $\Sigma_{{\rm s},2}$ &
  ILR  &
  CR  &
  \\
\hline
$13.1\kmskpc$& 0.15 \% & 47.8 $^{\circ}$ & 8.1 $^{\circ}$ & $6.6 \kpc$ & $26.4 \kpc$ & $0.13 \kpc$ & 24.9 \% & $4.1 \kpc$  & $17.6 \kpc$ &
\\
\hline
\hline
\\
 & & & & & \textbf{$\mathbf{\bsym{m}}$=3 spiral} & & & & & & & \\
 \hline
  $\Omega_{{\rm s},3}$ &
  $K_{{\rm s},3}$ &
  $\varphi_{{\rm s},3}$ &
  $p_{{\rm s},3}$ &
  $R_{{\rm s},3,{\rm min}}$ &
  $R_{{\rm s},3,{\rm max}}$ &
  $h_{{\rm s},3}$&
  $\Sigma_{{\rm s},3}$ &
  ILR  &
  CR  &
  \\
\hline
$16.4\kmskpc$ & 0.06 \% & 81.7 $^{\circ}$ & 13.7 $^{\circ}$ & $8.0 \kpc$ & $19.6 \kpc$ & $0.13 \kpc$ &  9.3 \% & $8.0 \kpc$  & $14.4 \kpc$ & \\

\hline
\end{tabular}
\end{table*}

\begin{figure*}[h]
\centering
\includegraphics[width=\hsize]{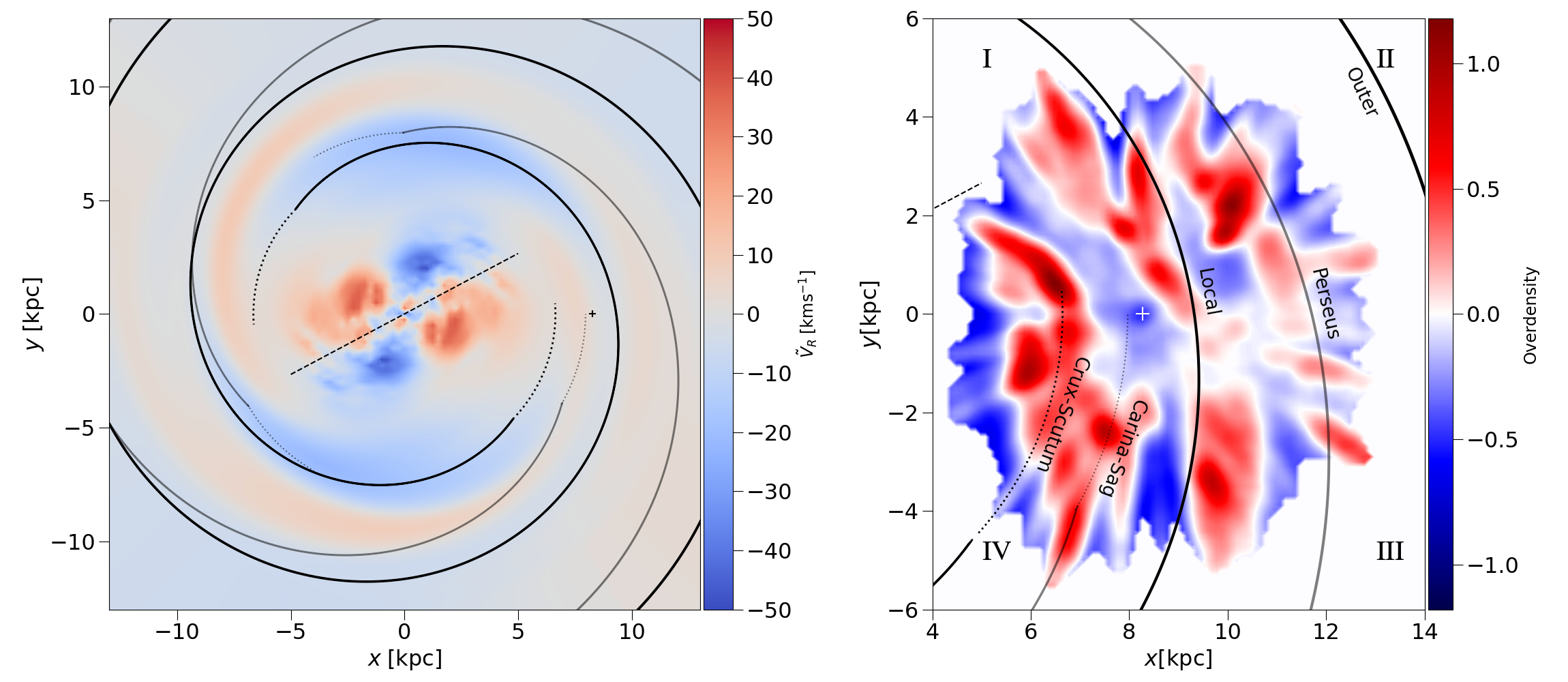}
\caption{Left panel: Median radial velocity $\tilde{V}_{R}$ for the fiducial non-axisymmetric model in the disk plane $(x,y)$ defined on $[-13, 13]\kpc \times [-13, 13]\kpc$ binned with $(250 \pc)^2$ bins. Right panel: Adaptation of panel B in figure~1 of \cite{Poggio2021} with over-densities of young upper main sequence stars in red tracing the position of the arms segments. Both panels display the bar position of our fiducial model with dashed black lines, the major $m=2$ (black) and minor $m=3$ (grey) spiral locations of our fiducial model as continuous and dotted black and gray lines, respectively. The continuous lines indicate the maximum overdensity of the model spirals as a function of radius from their (negative) potential minimum, down to the cutoff radius. The dotted line then traces an arc of circle at each cut-off radius, until the point where the spiral potential reaches zero. In both panels, the cross represents the Sun's position. The overdensity maps of young stars following \citet{Poggio2021} are generated from \texttt{https://github.com/epoggio/Spiral\_arms\_EDR3.git}.}
\label{fig:location}
\end{figure*}

\begin{figure*}[h]
\centering
\includegraphics[width=1.0\hsize]{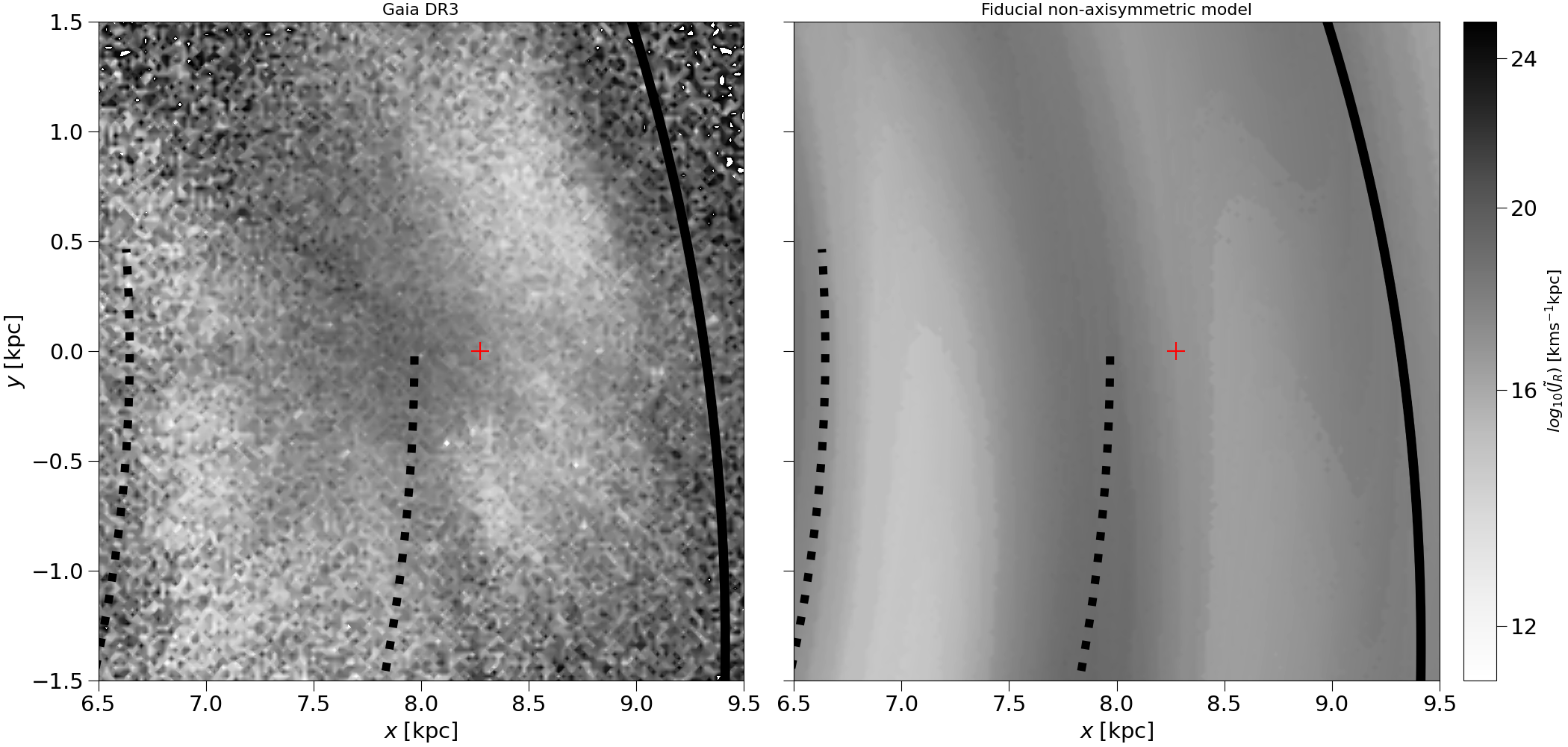}
\caption{The median $J_R$ as a function of position in the data of the extended Solar neighborhood (left panel) and in the model (right panel). The lines indicate the location of spiral arm segments in the model and the red cross indicates the Sun's position. 
}
\label{fig:JR}
\end{figure*}

\begin{figure*}[h]
\centering
\includegraphics[width=1.0\hsize]{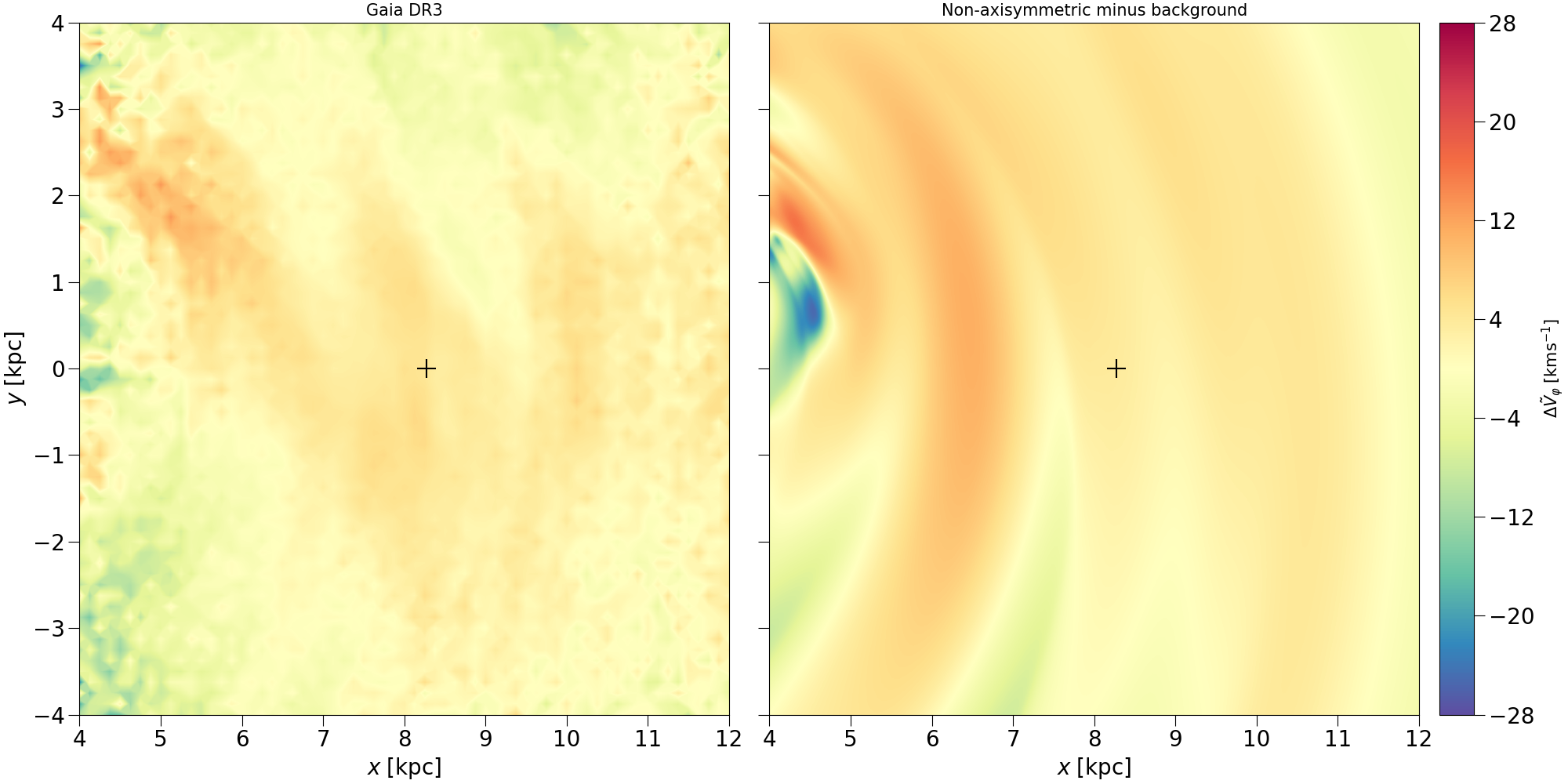}
\caption{
Left panel: difference of the median azimuthal velocity compared to its average value at fixed $R$ in the Gaia RVS data, $\Delta \tilde{V}_{\varphi} \equiv \tilde{V}_{\varphi}(x,y) - \tilde{V}_{\varphi}(R)$, in the $(x,y)$ plane defined on $[4, 12]~{\rm kpc} \times [-4, 4]~{\rm kpc}$, binned with $(125 \pc)^2$ bins. Right panel: difference between the median azimuthal velocity of the fiducial model and that of the background axisymmetric DF. In both panels, the black cross indicates the Sun's position. 
}
\label{fig:azimut}
\end{figure*}

With all the parametric components of the potential defined above, we are now in a position to launch our backward integrations to adjust the parameters to the data. As outlined here-above, the amplitude of the modes and the phase of the bar potential are fixed to values that fit well the dynamics of the bulge region. Only the pattern speed of the bar will now be adjusted to the location of resonant ridges in local velocity space, excluding the spiral arms from the model.

Another hyper-parameter to adjust and then fix is the (dummy) integration time, $T_{\rm int}$, within the backward integration context. This will not affect the location of ridges in local velocity space but will affect their apparent ``sharpness''. As in \citet{Dehnen2000}, we separate the total integration time into two equal-time phases of growth of the bar and plateau of its amplitude, with the following growth function:
\begin{equation}
    G_\mathrm{b}(t) = \begin{cases}
  1 & \text{if } -\frac{T_{\rm int}}{2} \leq t \leq 0,
  \\
  \frac{3}{16}{\cal T}^5 - \frac{5}{8}{\cal T}^3 + \frac{15}{16}{\cal T} + \frac{1}{2} & \text{if } -T_{\rm int} < t < -\frac{T_{\rm int}}{2},
\end{cases}
\end{equation}
where ${\cal T} \equiv  (4t+3T_{\rm int})/T_{\rm int}$. We choose to adjust those two parameters (pattern speed and dummy integration time) to the 1-D distribution of stars in the Solar neighborhood for azimuthal velocities within $90 \, \kms<V_{\varphi}<330 \, \kms$ at $V_R = 100 \, \kms$. This distribution is shown in Fig.~\ref{fig:bar_peaks}. The choice of analyzing the ridges at high $V_R$ prevents them from being ``contaminated'' by the additional effect of spiral arms since, as we shall see in the next subsection, these distort local velocity space mostly in the central regions of the velocity ellipsoid. This adjustment of the bar pattern speed is made in the Solar neighborhood, which is the most complete volume, so that peaks and valleys are not missing. 

Quantitatively, we compare the sum of the squares of the differences of the 1-D distribution of azimuthal velocities in each bin of $2 \, \kms$ between the Gaia RVS disk sample and the bar-only model. Only the location of the peaks matters here, so the DF renormalization is applied only in the small $V_R$ range considered in Fig.\ref{fig:bar_peaks}, instead of the DF normalization applied within the whole local velocity space in all other instances. We find the best match at $\Omega_{\rm b} = 37 \, \kms {\rm kpc}^{-1}$ for a total (dummy) integration time of 543~Myr corresponding to 3.2 rotations of the bar. Note however that the velocity peak that can be attributed to Bobylev moving group, or lower part of the Hercules bimodality (at $V_\varphi \sim 160 \kms$ in Fig.~\ref{fig:bar_peaks}), is not recovered, and is never so by a bar-only model that also reproduces the hat at large $V_\varphi$. Our best value of the pattern speed places the corotation radius of the bar at $R=6.6$~kpc and its OLR radius at $R=11$~kpc. 

In the first column of Fig.~\ref{fig:results}, we display the distribution of $(V_R,V_\varphi)$ velocities at the Solar position (setting the value to zero in pixels with no stars in the data within 300~pc from the Sun), of the local $(J_\varphi, J_R)$ action distribution, as well as the median Galactocentric radial velocity $\tilde{V}_R$ as a function of position within the Galactic plane. Remarkably, the local kinematic distribution corresponding to this {\it bar-only} model is already very similar to the observed one, without any additional contribution from spiral arms \citep[see also][for a less quantitative but similar conclusion]{Monari2019a}. The success of this bar model at producing so many features resembling the observed local kinematic distribution comes from the signatures of the Lindblad resonances of its multiple modes. We confirm this in the Appendix, where we provide a simple formula based on constant energy lines within the improved epicyclic formalism of \citet{Dehnen1999b} in order to evaluate the approximate location of the signature of each bar resonance in local velocity space. At $V_R = 100 \, \kms$, these approximate locations of the bar resonances are also indicated as small dashes on top of Fig.~\ref{fig:bar_peaks}. However, as it also appears clearly in the third row of Fig.~\ref{fig:results}, the bar-only model produces a dipolar structure of median radial velocities within the plane, far from the observed one. This implies that other dynamical ingredients are required to reproduce this median velocity field, which will be the topic of the following subsection. Another clear defect of the bar-only model, locally, is that the Sirius moving group does not stand out in local velocity space. Quantitatively, if one considers the density of stars within a strip of $V_\varphi$ between $250 \kms$ and $260\kms$ in local velocity space, and compares the value at $V_R=-12 \kms$ to that at $V_R=0 \kms$, one gets an increase of $\sim$25\% in the data at $V_R=-12 \kms$ (the Sirius peak), while one gets a {\it decrease} of 11\% in the bar-only model (almost identical to the axisymmetric case). This indicates that Sirius is likely caused by spiral arms. 

\subsection{Adding spiral arms}

\begin{figure*}[h]
\centering
\includegraphics[width=1.0\hsize]{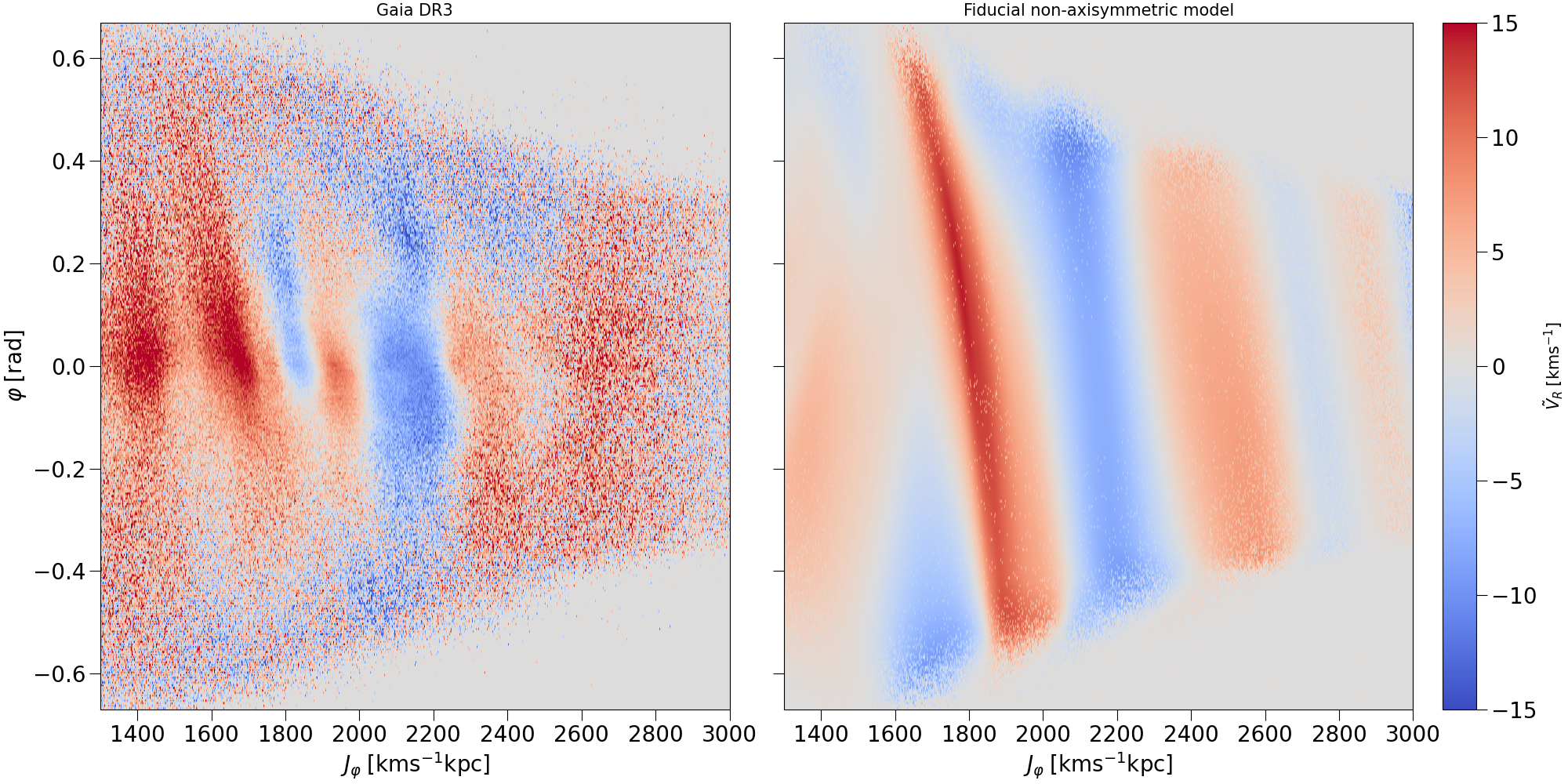}
\caption{
 Median radial velocity  in the $(\varphi,J_{\varphi})$ plane defined on $[-0.78,0.78]\, {\rm rad} \,\times \, [1000,3000] \, \kmskpc$, binned with bins of size $(0.005 \times 2.5 \; {{\rm rad} \, \times \,}\kmskpc)$. Left panel: the Gaia RVS disk sample. Right: the fiducial non-axisymmetric model. Points outside $5\kpc < x < 12\kpc$ and $-4\kpc < y < 4\kpc$ are excluded in both the data and model.
}
\label{fig:jphiphi}
\end{figure*}

Given the failure of the bar-only model to reproduce the median radial velocity field, the next step is to add non-axisymmetric modes corresponding to spiral arms.  
We start by adding a single mode on top of the bar-only model (i.e., with now fixed $\Omega_{\rm b} = 37 \, \kms {\rm kpc}^{-1}$), with multiplicity $m \in [1, 2, 3, 4]$. We fix the scale height to be the same as that of the gas component of the background potential, $h_{\mathrm{s},m}=130 \, \pc$,  the outer cut-off to be the OLR of the spiral, and the inner cut-off to be the larger between the corotation radius of the bar and the ILR of the spiral (so that the spiral lives between its ILR and OLR but does not penetrate within the corotation of the bar). The growth function $G_{\rm s, m}(t)$ has the same form as the bar, and we fix the integration time to exactly one full rotation of the spiral arm mode. In many other attempts, even allowing more than one rotation and a different growth time, the best candidates in the method that follows tend to converge close to the preferred values we found. 

The exploration of the whole parameter space with the backward integration method over a large portion of phase-space is computationally very costly, which led us to select the following strategy to fit the Galaxy model to the Gaia data. The fit is realized with the differential evolution method of \citet{Storn1997}, a global genetic optimization method implemented in the Python {\tt SciPy} library. This algorithm minimizes an objective function, set to be a weighted error function ${\cal L} = \sum_i  (\tilde{V}^{\rm model}_{R,i}-\tilde{V}^{\rm data}_{R,i})^2/\sigma_i^2$, comparing median radial velocities from model and data on a small selection of points $(x_i,y_i)$ with  weights $1/\sigma_i$. The observed median radial velocities $\tilde{V}^{\rm data}_{R,i}$ are calculated within bins of size $250 \, \pc$ around the selected point $(x_i,y_i)$, whilst the model median radial velocities are the median of the $V_{R}$ distribution at the selected point, i.e. the model DF values in the $(V_{R}, V_{\varphi})$ plane integrated over $V_{\varphi}$. The choice of the selected points and their respective weights is a delicate one. The number of points must be limited, in order to limit the computation time, but this also means that they must be chosen at `strategic' positions and not simply on a uniform grid. Moreover, simply weighting them by the number of stars in the data would give too much weight to the Solar vicinity over the entire area of the fit. The first point to which we nevertheless still give the highest weight, $1/\sigma_0$, is the Solar position $(x_0,y_0)$. We then need to choose points which are representative of the variations of the (positive and negative) values of the median radial velocity all over the plane. Adding spiral arms invariably runs the risk of not preserving the roughly correct radial velocity gradient from the bar in the region around $(x_1,y_1)=(7.0\kpc, 3.5\kpc)$ and $(x_2,y_2)=(7.0\kpc, 1.0\kpc)$, but it is needed to change the sign of $\tilde{V}_R$ at $(x_3,y_3)=(9.0\kpc,0.0\kpc)$. These are our three second-most important points, all with $\sigma_i=2\sigma_0$. We then choose two pairs of points along constant $y$ axes that encapsulate the positive-negative variations of the median radial velocity field, $(x_4,y_4)=(6.5\kpc, 0.0\kpc)$, $(x_5,y_5)=(10.0\kpc, 0.0\kpc)$, $(x_6,y_6)=(7.0\kpc, -3.0\kpc)$, and $(x_7,y_7)=(10.0\kpc, -3.0\kpc)$, with $\sigma_i=3\sigma_0$. In order to capture the clear spiral feature at the bottom-left of the plane, we also add two points, $(x_8,y_8)=(6.0\kpc,-2.5\kpc)$ and $(x_9,y_9)=(7.5\kpc, -2.5\kpc)$, with $\sigma_i=5\sigma_0$. We finally impose a constraint in the outer disk, $(x_{10},y_{10})=(11.5\kpc, 1.0\kpc)$ and $(x_{11},y_{11})=(12.0\kpc, 0.0\kpc)$, also with $\sigma_i=5\sigma_0$. These are the essential points of our fit. We add on top of this a set of low-weight points that will merely help guiding the fit, $(x_{12},y_{12})=(9.0\kpc, -3.0\kpc)$, $(x_{13},y_{13})=(9.0\kpc, 3.5\kpc)$, $(x_{14},y_{14})=(10.0\kpc, 3.5\kpc)$, and $(x_{15},y_{15})=(12.0\kpc, 3.5\kpc)$, all with $\sigma_i=100\sigma_0$. All the selected points are indicated as circles in the bottom-middle panel of Fig.~\ref{fig:results}. This selection of points and their weights hereafter plays the role of a prior on what the most important regions of configuration space are. 

For our genetic algorithm, let us now define our population of candidate solutions in parameter space as ${\mathbf{a}_{i,g}}$, with $1 \leq i \leq n$ and $1 \leq g \leq N$. This means we will consider $n$ candidates at each generation, for $N$ generations. In practice, a first generation of candidate solutions is created by picking stochastically many candidate parameters across parameter space by a Latin hypercube sampling, all while trying to cover most of the parameter space within the bounds specified hereafter. This population is then mutated, candidate by candidate, iteratively, thereby establishing a new generation at each iteration. At each generation $g$, the mutation of each candidate $\mathbf{a}_{i,g}$ is applied according to the ``best1bin" strategy with the following steps: 
\begin{itemize}
    \item Select the best parameters candidate (the one minimizing the weighted error function at current generation), $\mathbf{a}_{best,g}$.
    \item To mutate each candidate $\mathbf{a}_{i,g}$, randomly select two other parameters vector candidates, $\mathbf{a}_{j,g}$ and $\mathbf{a}_{k,g}$.
    \item Take a fixed multiplication factor (mutation factor ${\cal M}$) of their difference in parameters, ${\cal M} \, (\mathbf{a}_{j,g} - \mathbf{a}_{k,g})$
    to get a vector $\mathbf{v}_i = \mathbf{a}_{best,g} + {\cal M} \, (\mathbf{a}_{j,g} - \mathbf{a}_{k,g})$
    \item The new trial vector $\mathbf{a}_{i,g+1}$ is then built component by component by assigning the value of each parameter either from $\mathbf{v}_i$ or from $\mathbf{a}_{i,g}$ according to if a realization of the binomial function between 0 and 1 is smaller or greater than a chosen recombination value ${\cal C}$, respectively. 
    \item Compute the weighted error function for the trial vector $\mathbf{a}_{i,g+1}$: if it performs better in terms of the objective function, it replaces the original candidate in the next generation, otherwise the initial candidate $\mathbf{a}_{i,g}$ remains the same at generation $g+1$.
    \item The convergence criteria are met when the standard deviation of the population objective function values at a given generation is smaller than 1\% of the mean objective function value of all candidates in the population in that generation. The final $\mathbf{a}_{best,g=N}$ candidate is kept.
\end{itemize}
We kept the standard values of the algorithm hyper-parameters, notably population size $n$ (15 times the number of parameters), the recombination value ${\cal C}=0.7$, and the mutation factor ${\cal M}$, a random variable with values between 0.5 and 1. This method was chosen since it is extremely efficient at converging efficiently over a large parameter space. the selection of points and their weights plays the role of a prior on what the most important regions of configuration space are. However, contrary to a classical Bayesian method no posterior or well-defined error bars can be given. Therefore, we are not in a position to provide error bars, and we cannot exclude that our best candidate models found hereafter may be local minima in parameter space. Further improvements of the present work should address this question together with taking into account a Gaia selection function \citep[e.g.,][]{CastroGinard23}.
 
We first attempted to fit only one spiral arms mode, allowing pitch angles to vary between $6^{\circ}$ and $30^{\circ}$, the phase to vary all over $360^{\circ}$, the potential amplitude to vary from zero up to $0.2$\%, and the pattern speed from $10 \, \kmskpc$ up to the pattern speed of the bar: the mode $m=2$ performed the best in terms of the objective function among $m \in [1, 2, 3, 4]$, with a pattern speed of $13 \, \kmskpc$. This is the main result of our search, which we will now seek to refine. Indeed, this preferred single mode model clearly produces a distorted local velocity space, especially a very distorted Sirius-like moving group compared to observations. This is not entirely surprising, as local velocity space has not been used to constrain the fit. We then modify the objective function ${\cal L}$ with a local constraint, as follows: 
   $ {\cal L}'={\cal L}+\sum_{i=1}^{i=2}(\Delta_i^{\rm model}-\Delta_i^{\rm data})^2/\sigma_\Delta^2$, 
where $\Delta_i$ is the location of the 1-D $V_R$ distribution peak at $V_{\varphi} \, = \, 250 \, {\rm kms}^{-1}$ ($i=1$) and at $V_{\varphi} \, = \, 260 \, {\rm kms}^{-1}$ ($i=2$) at the Sun, and  $\sigma_\Delta=3\sigma_0$ in both cases. Using ${\cal L}'$ however still leads to a best candidate with a distorted Sirius moving group in local velocity space when considering a single $m=2$ mode. 

Then, in order to possibly improve over this model, we attempt a new fit that adds a second spiral mode with multiplicity $m=3$ or $m=4$, together with the the first one and the bar. We assume the $m=2$ spiral to have a range of pattern speeds $10\kmskpc < \Omega_{\mathrm{s},2} < 14\kmskpc$, close to the value found for the single mode fit which we aim to improve upon. To reduce the parameter space, the amplitude of the second higher mode -- whose pattern speed and pitch angle are allowed to vary from $10 \, \kmskpc$ up to the pattern speed of the bar and from $6^{\circ}$ to $30^{\circ}$ respectively -- is fixed with the equation proposed by \citet{Hamilton2024}, relating the respective amplitude of both modes to their pattern speed and pitch angle, namely as being inversely proportional to the product of their pattern speed squared with the tangent of their pitch angle (hence a higher amplitude for lower pattern speed and lower pitch angle). To further reduce parameter space, we impose that the sum of the local density contrasts for both spiral modes is smaller than 35\%, checking {\it a posteriori} that this limit will not be reached by our best candidate. To compute the surface density contrast of each mode, we take the ratio between the integrated surface density at the Sun of the axisymmetric baryonic component and the spiral arms surface density corresponding to the \citet{Cox2002} potential  \citep[see also][]{Monari2016b}.
In our analysis we found that the secondary $m=3$ spiral mode does complement the stronger mode better than the $m=4$ one in terms of the objective function. 
Adding this second $m=3$ mode allowed us to taper and regularize the signature of the Sirius moving group in local velocity space while improving slightly the median radial velocity map. To further polish the parameters of this best candidate found with our global optimization method, we then perform a fine search with a gradient descent in a narrow range of parameter space ($1 \kmskpc$ wide in pattern speed, $6^{\circ}$ wide in phase, $2^{\circ}$ wide in pitch angle and $0.04 \%$ wide in potential amplitude $K_{{\rm s},2}$) around our best candidate solution, with the Limited-memory Broyden–Fletcher–Goldfarb–Shanno (L-BFGS-B) algorithm implemented in {\tt Scipy}. The  solution thus found constitutes our fiducial model.

The final parameters of this fiducial model are presented in Table~\ref{table:non-axisymmetric}, while its local velocity and action space distribution, and median radial velocity map, are presented in the third column of Fig.~\ref{fig:results}. The improvement of the median radial velocity map compared to the bar-only model is striking, but there are also subtle improvements in local velocity space, in particular a better representation of moving groups close to the center of the velocity ellipsoid. For Sirius, if one reconsiders the density of stars within a strip of $V_\varphi$ between $250 \kms$ and $260\kms$, one now gets an {\it increase} of 5\% at $V_R=-12 \kms$ compared to $V_R=0 \kms$ in the model. This is still a significantly smaller peak than in the data ($\sim$25\%), which will require further investigations, but it is a significant improvement upon the {\it decrease} of 11\% in the bar-only model. We will now qualitatively compare the predictions of this fiducial model to other observables in the next Section. 

\begin{figure*}[h]
\centering
\includegraphics[width=1.0\hsize]{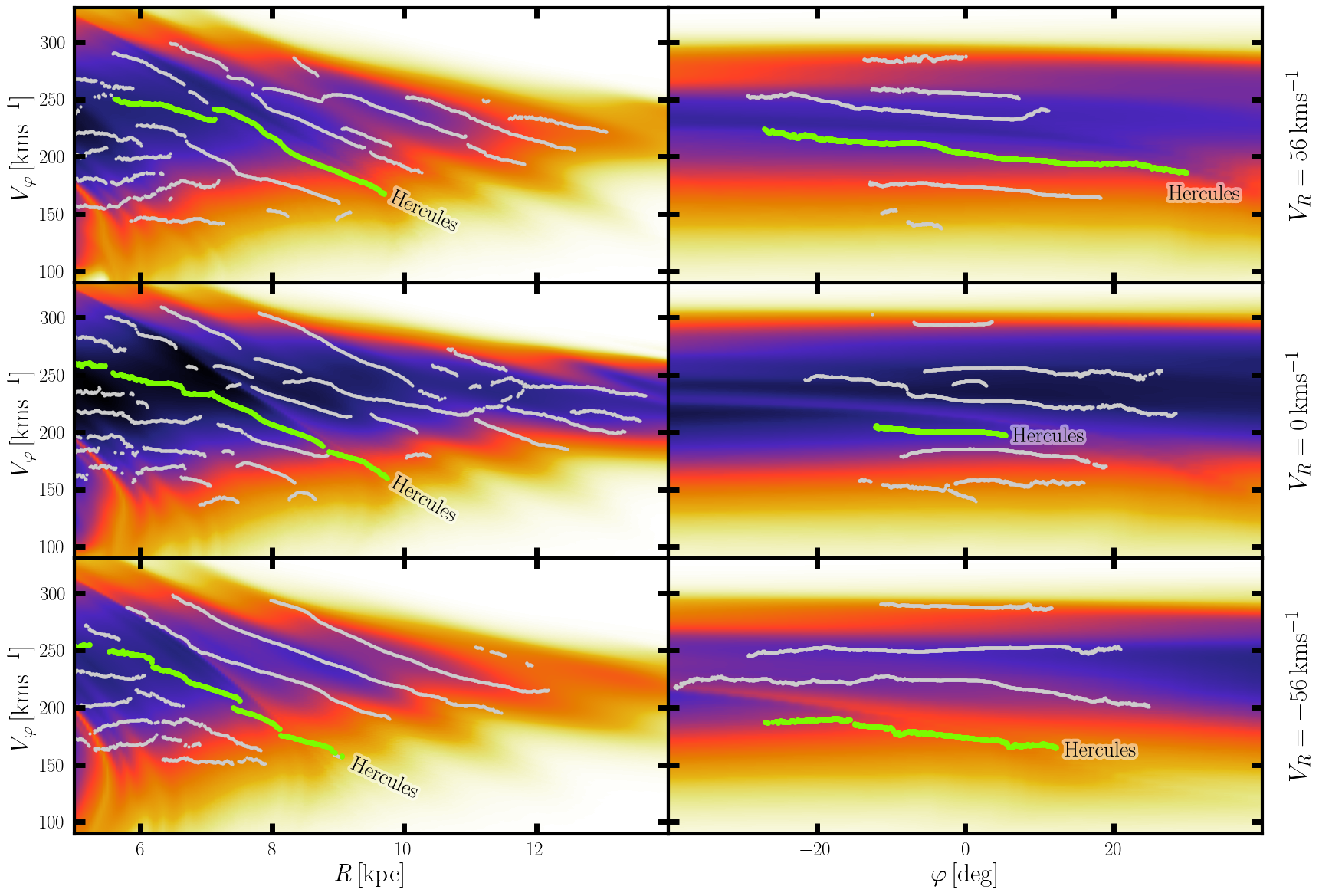}
\caption{
At fixed $V_R$ for each row ($V_R=56 \kms$, $V_R=0 \kms$ and $V_R=-56 \kms$, the white (and green) lines display the main ridges identified in Gaia DR3 {\it data} with the Wavelet Transform method developed in \cite{Bernet2022, Bernet2024}. Below these lines, we underlay the 2-D histogram distribution of the normalized fiducial {\it model}. Left: at fixed azimuth $\varphi=0^\circ$. Right: at fixed radius $R=R_0$. The Hercules ridge is denoted by a thicker green line.
}
\label{fig:ridges}
\end{figure*}

\begin{figure*}[h]
\centering
\includegraphics[width=1.0\hsize]{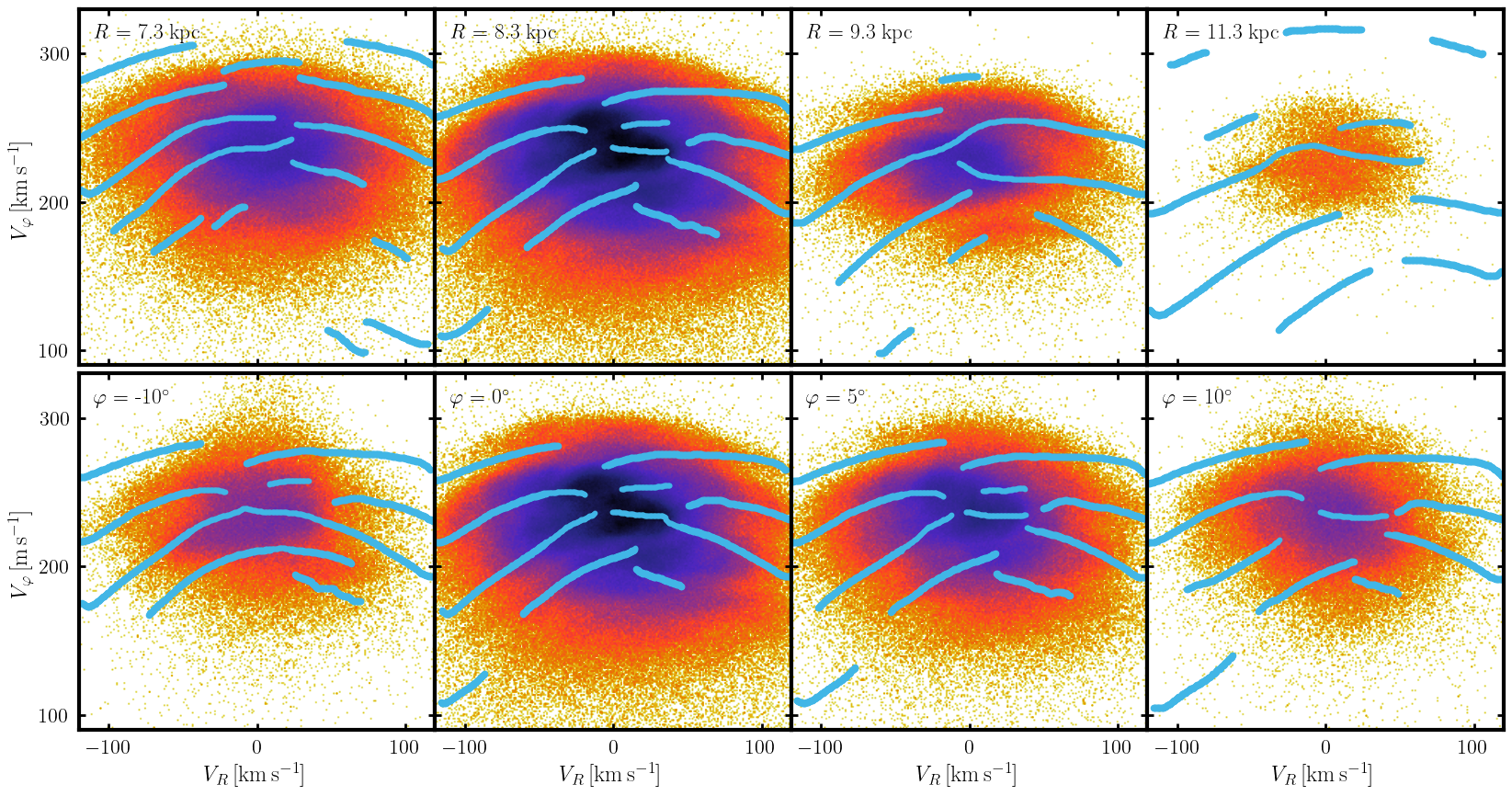}
\caption{
2-D histogram distribution of the Gaia RVS disk sample in the $(V_R,V_{\varphi})$ plane for stars within  $|\varphi| < 2.4^{\circ}$ and $|R-R_i| < 300 \, \pc$ at different radii $R_i \in [7.3, 8.3, 9.3, 11.3] \, \kpc$ (top row), and for stars within an annulus $|R-R_0|<300 \, \pc$ and $|\varphi - \varphi_j| < \arctan\left(\frac{300}{8275}\right)$ at different azimuthal angles $\varphi_j \in [-10^{\circ}, 0^{\circ}, 5^{\circ}, 10^{\circ}]$  (bottom row). The colored lines show the peaks identified in our {\it model} with the method of \citet{Bernet2022,Bernet2024}.
}
\label{fig:mgR}
\end{figure*}

\section{Predictions and applications of the model} \label{sec:appli}

The fiducial model presented in Table~\ref{table:non-axisymmetric} has been adjusted to data with zero prior on the spiral arms locations. One can now check how well the model performs in recovering the position of known spiral arms of the Galaxy, as well as how it fares in reproducing other observables, such as the azimuthal velocity field or the detailed locations of moving groups across the disk. 

\subsection{Locations of spiral arms}
On Fig.~5, we present the global radial velocity map predicted by our fiducial model, together with the location of the bar and of the maximum density of the two dynamically-fitted spiral arm modes\footnote{The continuous line for each spiral mode in Fig.~5 corresponds to the minimum of the spiral potential, and hence to the maximum spiral density, down to the cut-off radius. The correspondence between density and potential is not well defined on the circle corresponding to the cut-off radius, but choosing a smooth function $\sim (1+$tanh$)/2$ instead of a step function for $H_m(R)$ in Eq.~\ref{step} indeed does lead to a sharp density contrast on this circle until the point where the spiral potential reaches zero. This sharp density contrast is marked by the dotted lines in Fig.~5.}. We also compare the location of those arms to the over-densities of young upper main sequence stars identified in \citet{Poggio2021}. As it appears clearly in this figure, the strongest $m=2$ mode nicely matches the location of the Crux-Scutum arm close to the Galactic bar (although this arm location is also often labeled as a continuation of the Carina-Sagittarius arm), of the Local arm close to the Sun, and of the Outer arm. However, the distribution of young stars is a consequence of the distribution of the gas, while what we trace is the potential. Therefore, it is most useful to note that our results also appear in line with the findings of \citet{Widmark2024}, who found the Local arm to be a strong local over-density, with a contrast density of roughly 20\%, close to the local over-density of 24.9\% within our model. Since the pattern speed of the $m=2$ spiral is smaller than that of the bar, this could be interpreted as a recent disconnection (52.5 Myr ago) from the bar in the Crux-Scutum region, in accordance with the findings of \citet{Vislosky2024}. On the other hand, the weaker $m=3$ spiral nicely matches the location of the Carina-Sagittarius and Perseus arms. It is remarkable that a purely dynamical fit mostly recovers the location of known spiral arm over-densities within the disk.

Another interesting quantity to compare our model predictions with is the median $J_R$ as a function of position in the disk. Indeed, \citet{Palicio2023} identified spiral arm structures in the disk from the median $\tilde{J}_R$ values as a function of position. We reproduce such a map from the Gaia RVS data within the extended Solar neighborhood, in Fig.~\ref{fig:JR}, and overlay the location of the spiral arms from our fiducial model. We also compute the median axisymmetric $\tilde{J}_R$ values from our model, starting from the same grid of velocities as before at each location in the plane, then computing the corresponding radial actions with {\tt AGAMA} (in the axisymmetric background potential), and computing the median from the DF values. Again, the {\it a posteriori} qualitative agreement with the data is remarkable. Note that the increase in median axisymmetric $\tilde{J}_R$ is positively associated to the presence of spiral arms in our model, in accordance with the findings of \citet{Debattista2024} when considering instantaneous axisymmetric actions. In $N$-body simulations, one typically needs to average actions over a long-enough timescale to even better track the spirals for {\it low} values \citep{Debattista2024} of the median time-averaged radial action. In our case, the important takeaway is the {\it a posteriori} qualitative agreement between the data and model for the instantaneous $\tilde{J}_R$, without having used this quantity in the fitting procedure.

\subsection{Azimuthal velocity field}

An interesting quantity to look at in principle is the variation of the median azimuthal velocity at fixed Galactocentric radius, as this is also a clear signature of the non-axisymmetry of the potential. To avoid being dominated by the background DF and axisymmetric potential, one can plot from the data the value $\Delta \tilde{V}_{\varphi} \equiv \tilde{V}_{\varphi}(x,y) - \tilde{V}_{\varphi}(R)$, in the $(x,y)$ plane. This is shown in the left panel of Fig.~\ref{fig:azimut}. One drawback of showing this quantity is that the azimuthal concatenation at fixed $R$ can only be done in the region where data are available, which is why it was not obvious how to implement such a quantity as a target for the fit itself. From our fiducial model, on the other hand, one can directly subtract from the median radial velocity at each position the median radial velocity obtained from the background DF at the same location. Similar trends to the data can be seen in the model, although the two quantities are not straightforward to compare quantitatively. 

\subsection{Median radial velocity in the azimuth-angular momentum plane}

An interesting projection of Gaia data \citep[see, e.g.,][]{Friske2019,Monari2019b,Trick2021,Chiba2021a} is the structure of median (or mean) radial velocity in the azimuth-angular momentum plane. In Fig.~\ref{fig:jphiphi}, we display the median radial velocity in the $(J_{\varphi},\varphi)$  plane for stars within a box $[1300,3000] \, \kms \kpc \,\times \, [-0.67,0.67]\, {\rm rad}$, within $5 \, {\rm kpc} < x < 12 \, {\rm kpc}$ and within $-4 \, {\rm kpc} < y < 4 \, {\rm kpc}$. To compute the median values in the model, we first fix an azimuth $\varphi_{i}$ every 0.01~rad, then consider radii $R_{j}$ spaced $10$~pc from one another. Next, for each point we compute the DF with the backward integration method for different velocities $V_R$ and $V_{\varphi,n} = \frac{J_{\varphi,n}}{R_{j}}$. We then fix $J_{\varphi,n}$ and sum the values of the DF for all radii $R_{j}$, and we compute the median radial velocity for each $(\varphi_{i}, J_{\varphi, n})$. The qualitative agreement with the data is acceptable, although one can note that the signatures become weak at low $J_\varphi$ in the model. This can be related to the fact that our non self-consistent procedure is not particularly reliable in the very inner disk close to the bar region. It could also reveal that our constant pattern speed bar is not enough to explain the richness of the data within this plane \citep{Chiba2021a}, that we are missing the effect of vertical perturbations \citep[e.g.][]{Laporte2019,Laporte2020}, as well as accreted prograde structures, although all this would require further investigations. 

\subsection{Moving groups across the disk}

In \citet{Bernet2022,Bernet2024}, a methodology was developed to perform a blind search for moving groups in Gaia data across the whole disk, based on the execution of a Wavelet Transform in independent small volumes of the disk followed by a grouping into global structures with the Breadth-first search algorithm from Graph Theory. Fixing a given $V_R$, one can then for instance look at the evolution of the location of moving groups in the $(R,V_\varphi)$ plane at the azimuth of the Sun, or in the $(\varphi,V_\varphi)$ plane at the radius of the Sun.  In Fig.~\ref{fig:ridges}, we overlay the structures found in Gaia DR3 on top of the density from our model. The azimuthal distribution of moving groups (right panels of Fig.~\ref{fig:ridges}) is well in line with the slopes from our model at the Solar radius, while the radial distribution at the Solar azimuth (left panels of Fig.~\ref{fig:ridges}) also appears globally in line with the data apart from the low $V_\varphi \, \leq 200 \, \kms$ region for small $R \, \leq 6.5 \, {\rm kpc}$ (the ridges of the model having a much too high slope in that region of phase-space), where the bar self-gravity is probably having a non-negligible effect on the data. 

Interestingly, in the model, the Hercules moving group at the Solar radius appears to result from the merging of two ridges at smaller radii, seen as dark regions in Fig.~\ref{fig:ridges} in the underlying density of the model, one with a slope compatible with the observed radial gradient of the Hercules moving group, essentially caused by the bar, and another one with a larger slope, mostly due to spiral arms. This is especially clear at positive $V_R$, where the two ridges are clearly separated at $R<7$~kpc in the model, while this separation appears to leave a similar signature within the data too.  At $V_R=0$, the split can also be seen, although it also merges with the inner continuation of the Hyades moving group. At negative $V_R$, the agreement is less good, though in the region where Hercules is expected to dominate less: the second ridge of Hercules overlaps with the Hyades ridge at $R \sim 7 \,$kpc in the model, whilst in the data this is only seen as a small upward bend in the Hyades ridge, corresponding to the merging of the ridges in the model. This second Hercules ridge is clearly an effect of spiral arms, while the major Hercules one is produced by the bar alone. The joint effect of the multiple bar modes in the present model, together with the axisymmetric background potential used, might explain the differences with \citet{Bernet2024}.

Conversely, we apply the method of \citet{Bernet2022,Bernet2024} on the {\it model} and overlay in Fig.~\ref{fig:mgR} the detected groups on top of Gaia data at different radii at the azimuth of the Sun, and at different azimuths at the Solar radius. Visually, some features are stikingly similar in the model and data. An interesting point to note is that, even though not clearly visible by eye, the model does seem to recover an overdensity at the Sun's position (bottom-right feature in the second panels from the left in Fig.~\ref{fig:mgR}) that can be identified with the location of the Bobylev/Hercules-2 bimodality of Hercules, although much less strongly than in the data. 

\subsection{Orbit of the Sun}

\begin{figure}[h]
\centering
\includegraphics[width=1.0\hsize]{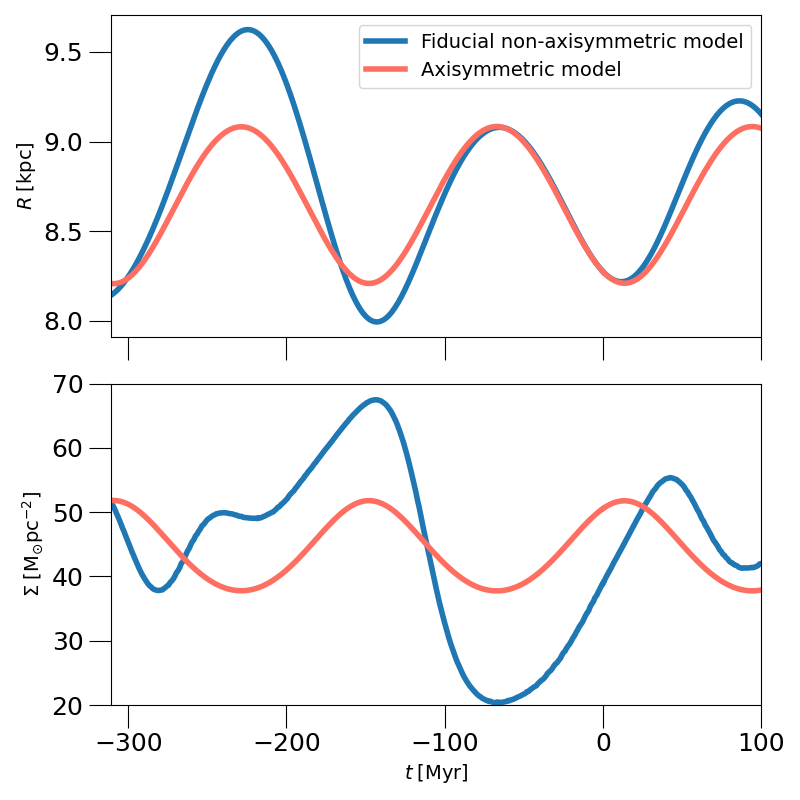}
\caption{Top panel: Galactocentric radius of the Sun as a function of time in the background axisymmetric (orange) and fiducial non-axisymmetric (blue) models. Bottom panel: Evolution of the surface density at the Sun's position with time.}
\label{fig:sun_orbit}
\end{figure}

As an example of application of our model, we propose to compare the in-plane orbit of the Sun in the background axisymmetric model to that in our fiducial non-axisymmetric one. The result is displayed in Fig.~\ref{fig:sun_orbit}. It is mostly illustrative, and should not be over-interpreted given that the vertical motions are neglected. In the axisymmetric case, the radial period is 161.5~Myr, and the Sun is now close to reaching its pericenter. The time between the last pericentric passage and the one that we are about to reach is a bit smaller in the non-axisymmetric model, namely 154.5~Myr. The previous pericentric passage, which happened a bit later in the non-axisymmetric model, was also closer to the Galactic center than in the axisymmetric case. The last apocenter was very similar in both models, but the next-to last one was further away in the outer Galaxy in the non-axisymmetric case, for which radial amplitudes are typically larger. If we look at the evolution of the surface density at the Sun's position with time, the picture becomes more complicated. The time between the last surface density maximum and the one that we are about to reach (i.e., still 161.5~Myr in the axisymmetric case) is a bit larger in the non-axisymmetric case, namely 185~Myr, because we will be temporarily following the Local arm overdensity on our journey back to the outer disk. Also, when looking back at $t \sim -250$~Myr, the apocenter that corresponds to a minimum in the surface density of the axisymmetric model does actually correspond to a local maximum in the non-axisymmetric case, because the Sun was also following a spiral arm at that time. This could have interesting consequences in studying cyclic sedimentation on Earth on long timescales \citep[e.g.,][]{Boulila}. Since spiral arms are generally expected to arise from a recurrent cycle of groove or edge modes, it is however impossible to trace back the Sun's orbit on longer timescales than a few 100~Myr, at least without resorting to detailed chemodynamical modeling of the evolution of the whole Galactic disk. 

\subsection{Orbits of young associations}

\begin{figure}[h]
\centering
\includegraphics[width=1.0\hsize]{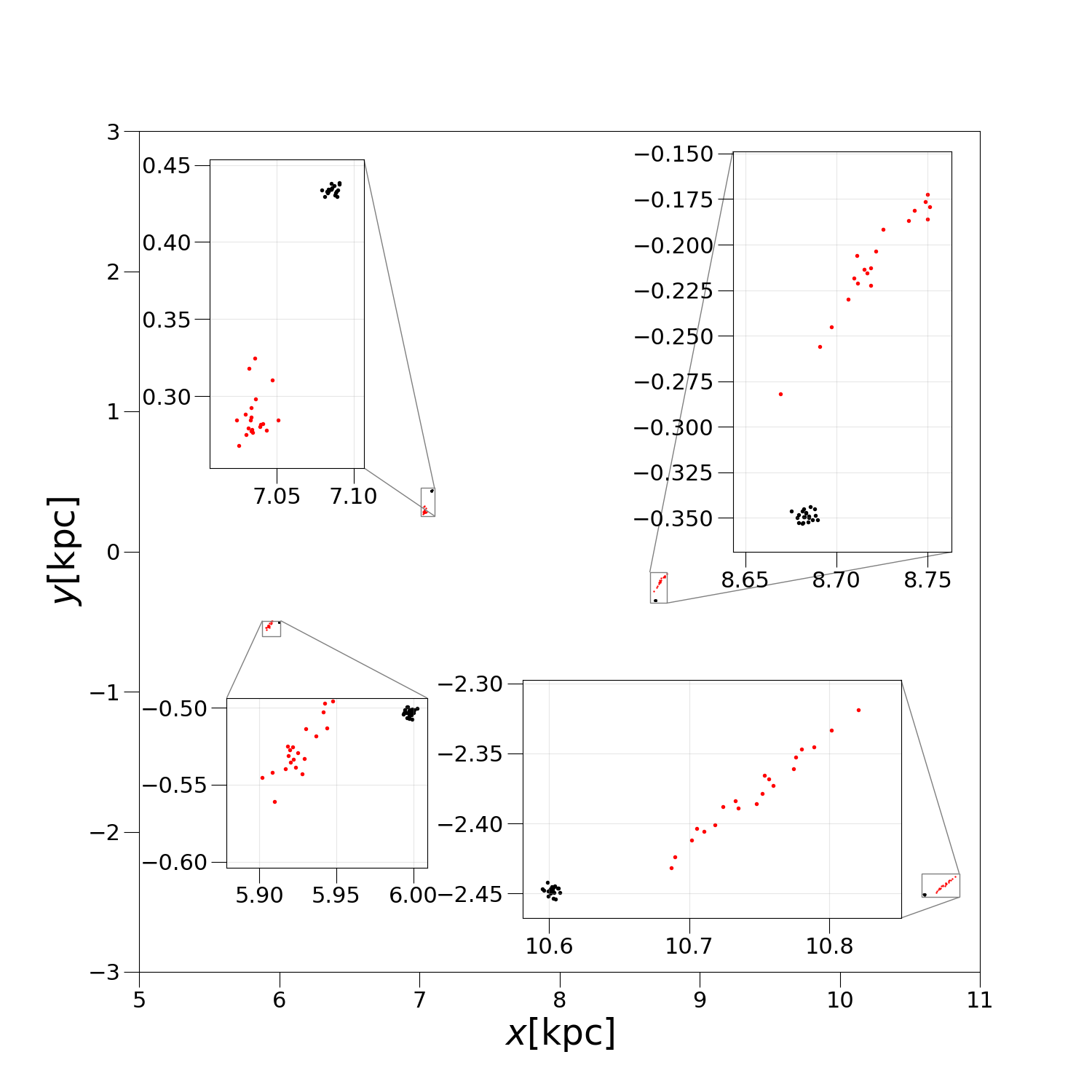}
\caption{
The implication of the non-axisymmetries on the orbits of young associations is illustrated by integrating four young associations for 50~Myr forward in the fiducial potential and then backward in time with both the erroneous axisymmetric model (red) and the fiducial one (black). The insets show zoom-ins around the regions of the associations.
}
\label{fig:young_associations}
\end{figure}

Young ($\sim 50$~Myr) stellar associations can typically be traced back to their original position by integrating their orbits backward in a given Galactic potential. To illustrate the importance of taking into account the non-axisymmetries of the potential for such a procedure, we integrate forward in time for 50~Myr four archetypal young stellar associations at different positions in the Galactic disk within our fiducial non-axisymmetric model. Each association is represented by 20 stars that are dispersed in velocity and space with Gaussians of standard deviation of $1\kms$ in $V_R$ and $V_{\varphi}$ around the circular velocity, and of $2\pc$ around the positions $(x,y) = (10.6\kpc,-2.4\kpc)$, $(8.7\kpc,-0.3\kpc)$, $(7.1\kpc,0.4\kpc)$, $(6.0\kpc, -0.5\kpc)$. We then integrate them backward in time both in the fiducial non-axisymmetric model and in the background axisymmetric model. The associations typically go back to a position that can be erroneous by more than 150~pc in the axisymmetric case, with an elongated shape very different from the true original configuration (Fig.~\ref{fig:young_associations}). 

%-------------------------------------- 

\section{Discussion and conclusions} \label{sec:concl}
We used the in-plane velocities of a sample of disk stars with six-dimensional phase-space information from Gaia-StarHorse in order to fit a Galactic potential model that includes a detailed parametric shape for the bar and spiral arms, using the backward integration method. The adjusted observable quantities were the median Galactocentric radial velocity, for a selection of points within the Galactic plane, complemented by additional constraints from local velocity space at the Sun's position. All the parameters of the fiducial non-axisymmetric potential are summarized in Table~\ref{table:non-axisymmetric}, within the background axisymmetric potential fixed in Table~\ref{table:axisymmetric}. It is remarkable that such a purely dynamical fit recovers many of the known locations of spiral arm over-densities detected from photometry within the disk (Fig.~\ref{fig:location}). These spiral arms can be interpreted as groove or edge modes such as those found in $N$-body simulations. In the Solar vicinity, we identify the Local arm as a strong gravitational perturbation, in accordance with independent probes of the local non-axisymmetric potential by \citet{Widmark2024}. We also recover the observed map of median radial actions in the extended Solar vicinity (Fig.~\ref{fig:JR}), as well as a good qualitative agreement with the detailed variation with radius and azimuth of moving groups identified in Gaia data (Fig.~\ref{fig:ridges}). The latter is truly remarkable since the model was not directly fit to these phase-space features. The only (locally) fitted moving group was Sirius, entirely absent from the bar-only model: we note that it is nevertheless still characterized by a weaker peak in the fiducial model than in the data. While our best candidate model may well be a local minimum in parameter space, the latter being particularly vast, especially if letting the parameters of the background axisymmetric density and DF vary too, we nevertheless argue that it can, for the time being, be used as a fiducial non-axisymmetric potential for the Galaxy, for instance in order to confidently integrate in-plane orbits. It can be compared to other recent Galactic potentials such as that of \citet{Hunter}. The code to generate local velocity space distributions as well as radial velocity maps is made public. 

This paper represents only a first quantitative step in the direction of establishing a detailed three-dimensional non-axisymmetric potential for the Milky Way. Future improvements of our present investigations will be to explore its three-dimensional consequences, first in forward test-particle integrations \citep[see, e.g.,][]{Faure2014,Monari2016b}, as both the bar \citep{Thomas23} and spiral \citep{Cox2002} potentials can readily be generalized to three dimensions. Let us also note that we have made the assumption that the spiral arms cannot live inside the corotation resonance of the bar ($R=6.6$~kpc in our fiducial model), an assumption that could potentially be partially lifted: we already checked that it does not affect much our best candidate model. However, a proper fit of these inner regions of the Galactic disk would require us to make the model self-consistent. The absence of self-consistency can become a serious issue in the innermost parts of the Galaxy, where the bar perturbation is itself the tracer. Some deficiencies of our model at low radii and low angular momenta have indeed already been identified in Fig.~\ref{fig:jphiphi} and Fig.~\ref{fig:ridges}. Although beyond the scope of the present work, future improvements of our model might rely on an adaptation of the self-consistent made-to-measure method in order to account for self-consistency. One should also point out that the background model (axisymmetric potential and DF) has not been fitted here, and could in principle also be adjusted to the data. This would increase the parameter space and might require us to use machine learning methods to efficiently explore the full parameter space. A parallel improvement will be to incorporate a detailed selection function in the fitting procedure \citep[e.g.,][]{CastroGinard23,Khanna2024}, compute a proper posterior (and error bars) on the best-fit parameters, while perhaps attempting to separate the stellar populations into distinct DFs. Finally, when moving to three dimensions, it is obvious that vertical perturbations of the disk, e.g. from the Sagittarius dwarf, will have to be included too, together with a possible time-variation of the bar pattern speed.

In summary, the fiducial model presented here, reproducing a larger amount of observables than ever before, does represent a significant advance in our understanding of the non-axisymmetric structure of the Milky Way disk. However, it is important to recognize its limitations and to continue improving it in order to obtain an even more accurate representation of our Galaxy.

\section*{Code availability}

The ``Spiral and Bar Backward Integration" ({\tt SPIBACK}) {\tt PyTorch}-based code to generate local velocity space distributions as well as median radial velocity maps is publicly available at: \texttt{https://github.com/yrkhalil/SPIBACK}

\begin{acknowledgements}
    We thank the anonymous referee for constructive comments that helped improving the paper.
    YRK and BF thank Laurent Navoret for his support and advices throughout the project, as well as insightful comments from Leandro Beraldo e Silva, Pedro Alonso Palicio, Eugene Vasiliev, Jean-Baptiste Fouvry, Jorge Peñarrubia, Ronald Drimmel and Eloisa Poggio. This work was supported by the Interdisciplinary Thematic Institute IRMIA++, as part of the ITI 2021-2028 program of the University of Strasbourg, CNRS and Inserm, was supported by IdEx Unistra (ANR-10-IDEX-0002), and by SFRI-STRAT’US project (ANR-20-SFRI-0012) under the framework of the French Investments for the Future Program. BF, GM, AS, RI  acknowledge funding from the ANR grants ANR-20-CE31-0004 (MWDisc), ANR-19-CE31- 0017 (SEGAL) and ANR-18-CE31-0006 (GaDaMa). BF, AS and RI acknowledge funding from the European Research Council (ERC) under the European Unions Horizon 2020 research and innovation programme (grant agreement No. 834148). MB acknowledges funding from the University of Barcelona’s official doctoral program for the development of a R+D+i project under the PREDOCS-UB grant. TA acknowledges the grant RYC2018-025968-I funded by MCIN/AEI/10.13039/501100011033 and by ``ESF Investing in your future’’, the grants PID2021-125451NA-I00 and CNS2022-135232 funded by MICIU/AEI/10.13039/501100011033 and by ``ERDF A way of making Europe’’, by the ``European Union'' and by the ``European Union Next Generation EU/PRTR'' and the Institute of Cosmos Sciences University of Barcelona (ICCUB, Unidad de Excelencia ’Mar\'{\i}a de Maeztu’) through grant CEX2019-000918-M.

    This work has made use of data from the European Space Agency (ESA) mission {\it Gaia} (\url{https://www.cosmos.esa.int/gaia}), processed by the {\it Gaia} Data Processing and Analysis Consortium (DPAC, \url{https://www.cosmos.esa.int/web/gaia/dpac/consortium}). Funding for the DPAC has been provided by national institutions, in particular the institutions participating in the {\it Gaia} Multilateral Agreement. This research has made use of the SIMBAD database and of the VizieR catalogue access tool, operated at CDS, Strasbourg, France.
\end{acknowledgements}

\bibliographystyle{aa}
\bibliography{./biblio.bib}

\begin{appendix}

\section{A simple formula for approximately locating bar resonances in velocity space}
\label{analytic}

In order to evaluate analytically the location of the phase-space resonant zones due to a constant pattern speed bar perturbation, one should define those zones in terms of librating versus circulating orbits after canonically transforming to slow and fast action and angle variables \citep[e.g.][]{Monari2017a,Binney2020}. This is a relatively arduous task, and we propose here a less precise but much faster way to roughly estimate the location of resonant ridges in local velocity space, or more accurately the surfaces of phase-space where the resonant condition is fulfilled. This simple formula will be, by construction, axisymmetric, which means that one cannot use it to model the changes in the morphology of the resonant surfaces with azimuth. The formula relies on locally drawing constant energy lines in the $V_R$-$V_\varphi$ plane within the improved epicyclic formalism presented in \citet{Dehnen1999b}. For convenience, we reproduce here their equations~28 and 29 for the radial evolution of an orbit:
\begin{align}
\label{eq:dehnen_r}
R(\eta) &= R_E\left[1-e \, \mathrm{cos}(\eta)\right]^{\frac{\gamma}{2}},\\
\label{eq:dehnen_e}
e &= \sqrt{1-\left(\frac{J_{\varphi}}{J_{\varphi}^{\rm circ}}\right)^2},
\end{align}
\noindent where $J_{\varphi}^{\rm circ}$ and $R_E$ are, respectively, the angular momentum and radius of a circular orbit of the same energy, $\gamma=\gamma(R_E) =2\Omega(R_E)/\kappa(R_E)$ is the ratio of the circular frequency $\Omega$ over the radial epicyclic frequency $\kappa$ as usual, and $e$ is the eccentricity of the orbit. The phase in radius evolving with time $t$, namely $\eta(t)$, is the parameter that represents the position along the path of the orbit, and is defined as a function of the eccentricity and the radial epicyclic frequency \citep[see Eq.~28d of][]{Dehnen1999b}. For simplicity, we consider that $\eta \sim \kappa t$, which in turn means that the radial velocity can be calculated as:
\begin{equation}\label{eq:dehnen_vr}
    V_R = \frac{dR}{d\eta}\frac{d\eta}{dt} \sim \kappa \frac{\gamma}{2}R(\eta)\frac{e \, \mathrm{sin}(\eta)}{1-e \, \mathrm{cos}(\eta)} = V_c(R_E)\frac{R}{R_E}\frac{e \, \mathrm{sin}(\eta)}{1-e \, \mathrm{cos}(\eta)}.
\end{equation}
The azimuthal velocity, on the other hand, can be written as a function of $R$ simply by the conservation of angular momentum: $V_\varphi=J_{\varphi}/R(\eta)$. Using Eq.~\ref{eq:dehnen_r} and Eq.~\ref{eq:dehnen_e} , one then gets a closed form equation that relates the changes of $V_R$ along the orbit to those of $R$, $V_\varphi$ and the circular velocity curve: 
\begin{equation}\label{eq:vr_without_eta}
    V_R = V_c(R_E)\left(\frac{R}{R_E}\right)^{\frac{\gamma-2}{\gamma}}\sqrt{2\left(\frac{R}{R_E}\right)^{\frac{2}{\gamma}}-\left(\frac{R}{R_E}\right)^{\frac{4}{\gamma}}-\left(\frac{R V_\varphi}{J_{\varphi}^{\rm circ}}\right)^2} .
\end{equation}
All that is left is to impose the constraint that stars be on a surface where the resonant condition is met. Under the assumption that frequencies are functions solely of the energy, the resonant condition is preserved as long as stars share the same $J_{\varphi}^{\rm circ}$ and $R_E$, and as longs as these are equal to the radius and angular momentum where the circular orbit is $l:m$ resonant. We refer to these as  $R_{\rm res}$, such that $l \, \kappa(R_{\rm res}) + m \, [\Omega(R_{\rm res})-\Omega_b]=0$, and as $J_{\varphi}^{\rm res} \equiv R_{\rm res} V_c(R_{\rm res})$, respectively. The final equation for the surfaces of resonant condition is then:
\begin{equation}\label{eq:dehnen_vphi_resonant}
    V_\varphi^{\rm res}(R,\,V_R) = \frac{J_{\varphi}^{\rm res}}{R}\sqrt{2\left(\frac{R}{R_{\rm res}}\right)^{\frac{2}{\gamma}}-\left(\frac{R}{R_{\rm res}}\right)^{\frac{4}{\gamma}}\left[1+\left(\frac{R_{\rm res}}{R}\frac{V_R}{V_c(R_{\rm res})}\right)^2\right]}.
\end{equation}
\noindent where we reordered the terms to express $V_\varphi$ as a function of the other phase-space variables.

\begin{figure}[h]
\centering
\includegraphics[width=1.0\hsize]{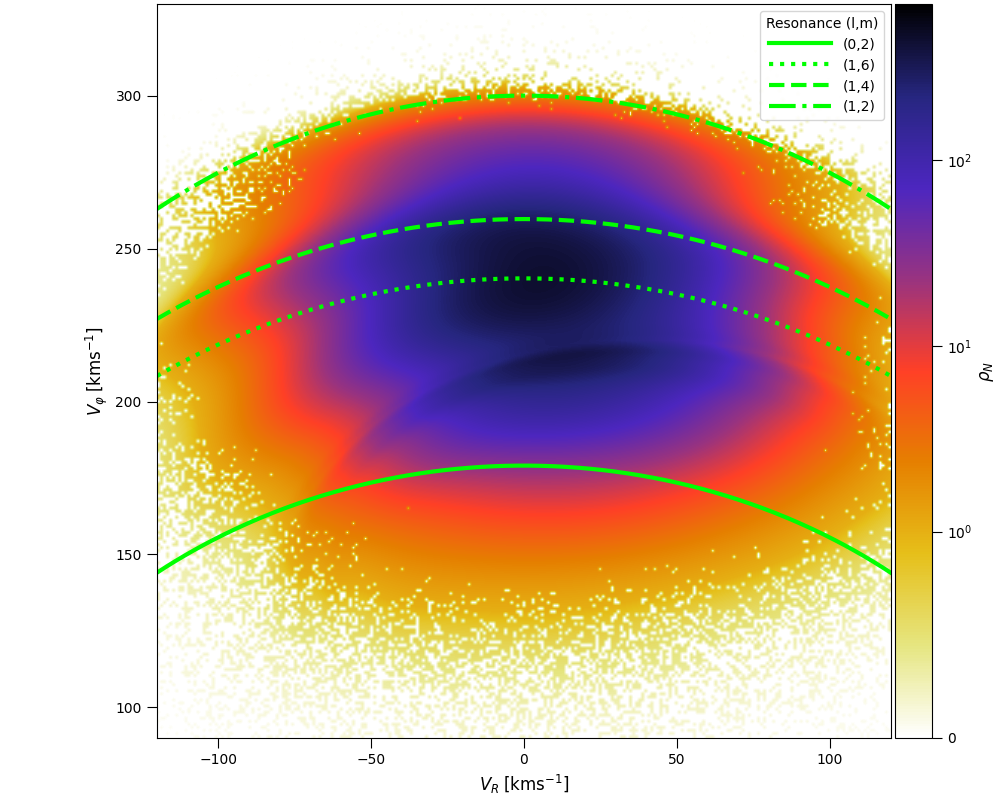}
\caption{Resonant lines $V_\varphi^{\rm res}$ as a function of $V_R$ from Eq.~\ref{eq:dehnen_vphi_resonant} at $R=R_0$, computed with the potential of Table~\ref{table:axisymmetric}, for a bar pattern speed of $37$~km~s$^{-1}$~kpc$^{-1}$. The lines are overlaid on top of the local velocity space density produced by our preferred bar-only model at the Sun's position.}
\label{fig:resonances_formula}
\end{figure}

The application of Eq.~\ref{eq:dehnen_vphi_resonant} to the $V_R$-$V_\varphi$ velocity plane produces for each resonance a curve that is almost identical to Eq.~9 from \citet{Dehnen2000}, and is equivalent to drawing lines of constant energy at a given configuration space point, because it is written under the assumption that the orbital frequencies are pure functions of the orbital energy. As can be seen in Fig.~\ref{fig:resonances_formula}, in our best bar model, this formula identifies extremely well the peak of the overdensities produced at $R=R_0$ by the $6:1$, $4:1$ and $2:1$ outer resonances of the bar. For the $(l,m)=(0,2)$ corotation resonance, the formula fails, probably because the eccentricty becomes too large, and it rather approximately identifies the lower bound of the resonant zone.

\end{appendix}

\end{document}